\documentclass[11pt]{article}

\usepackage[utf8]{inputenc}
\usepackage[T1]{fontenc}
\usepackage[american]{babel}

\usepackage{arxiv}

\usepackage{palatino}

\usepackage{graphicx}
\graphicspath{{figures/}}
\usepackage{pgfplots}
\pgfplotsset{compat=1.18}
\usepgfplotslibrary{groupplots}
\usetikzlibrary{backgrounds,patterns}

\usepackage{amsmath, amsthm, mathtools}
\usepackage{amssymb}
\usepackage{bm}
\usepackage{booktabs}
\usepackage{multirow}
\usepackage{array}
\usepackage{tabularx}
\usepackage{caption}
\usepackage{subcaption}
\usepackage{float}

\usepackage{hyperref}
\usepackage[capitalise,noabbrev]{cleveref}
\usepackage{url}

\usepackage[square, sort&compress, numbers]{natbib}

\usepackage{xcolor}
\usepackage{colortbl}
\definecolor{sonaNavy}{HTML}{263340}
\definecolor{sonaServingTint}{HTML}{F2F4F7}
\definecolor{sonaBrown}{HTML}{8C3F28}
\definecolor{sonaTrainingTint}{HTML}{F3EEEA}
\definecolor{linkInk}{RGB}{28,69,135}
\hypersetup{
  colorlinks=true,
  linkcolor=linkInk,
  citecolor=linkInk,
  urlcolor=linkInk,
}
\usepackage{enumitem}
\setlist{itemsep=2pt, parsep=0pt, topsep=4pt}
\usepackage{microtype}
\usepackage{xspace}

\usepackage{algorithm}
\usepackage{algpseudocode}

\usepackage{listings}
\renewenvironment{abstract}{\small\begin{quote}}{\end{quote}}

\newcommand{\Sona}{\textsc{Sona}\xspace}

\numberwithin{equation}{section}
\numberwithin{figure}{section}
\numberwithin{table}{section}

\newcommand{\NumBackboneLayers}{7}
\newcommand{\NumDecoderLayers}{2}
\newcommand{\NumFullHistoryLayers}{1}
\newcommand{\NumBridgeLayers}{1}
\newcommand{\NumHiddenDim}{1024}
\newcommand{\NumAttnHeads}{16}

\newcommand{\NumCodebookSize}{32{,}000}
\newcommand{\NumSemIDLevels}{3}
\newcommand{\NumScalingSmallCoreParams}{20M}
\newcommand{\NumScalingMediumCoreParams}{130M}
\newcommand{\NumScalingLargeCoreParams}{260M}
\newcommand{\NumScalingSmallTotalLearnableParams}{264M}
\newcommand{\NumScalingMediumTotalLearnableParams}{485M}
\newcommand{\NumScalingLargeTotalLearnableParams}{659M}

\newcommand{\NumDataScalingOneWeekWindow}{1 week}
\newcommand{\NumDataScalingTwoWeekWindow}{2 weeks}
\newcommand{\NumDataScalingFourWeekWindow}{4 weeks}
\newcommand{\NumDataScalingEightWeekWindow}{8 weeks}
\newcommand{\NumDataScalingOneWeekTrainLoss}{2.2735}
\newcommand{\NumDataScalingTwoWeekTrainLoss}{2.1061}
\newcommand{\NumDataScalingFourWeekTrainLoss}{2.0800}
\newcommand{\NumDataScalingEightWeekTrainLoss}{2.0427}
\newcommand{\NumDataScalingOneWeekRecallTen}{0.1798}
\newcommand{\NumDataScalingOneWeekRecallHundred}{0.4979}
\newcommand{\NumDataScalingOneWeekRecallThousand}{0.8333}
\newcommand{\NumDataScalingTwoWeekRecallTen}{0.2126}
\newcommand{\NumDataScalingTwoWeekRecallHundred}{0.5576}
\newcommand{\NumDataScalingTwoWeekRecallThousand}{0.8640}
\newcommand{\NumDataScalingFourWeekRecallTen}{0.2388}
\newcommand{\NumDataScalingFourWeekRecallHundred}{0.5993}
\newcommand{\NumDataScalingFourWeekRecallThousand}{0.8849}
\newcommand{\NumDataScalingEightWeekRecallTen}{0.2519}
\newcommand{\NumDataScalingEightWeekRecallHundred}{0.6178}
\newcommand{\NumDataScalingEightWeekRecallThousand}{0.8947}

\newcommand{\NumTokenizerEmbDim}{256}
\newcommand{\NumQwenSeqLen}{640}
\newcommand{\NumPopThreshold}{500}

\newcommand{\NumCompressMaxLen}{8192}
\newcommand{\NumScalingLongHistoryLength}{8k}

\newcommand{\NumCompressRecentLen}{2048}

\newcommand{\NumRewardRefreshCadence}{24 hours}
\newcommand{\NumRewardParams}{0.6B}

\newcommand{\NumRewardEncoderLayers}{10}
\newcommand{\NumRewardCrossLayers}{6}
\newcommand{\NumRewardHidden}{1024}
\newcommand{\NumRewardHeads}{16}
\newcommand{\NumRewardHistoryLength}{8k}
\newcommand{\NumRewardPretrainLength}{2k}

\newcommand{\NumRewardWeightedPairNoPretrain}{0.6153}
\newcommand{\NumRewardWeightedPairHistoryTwoKDepthOne}{0.6080}
\newcommand{\NumRewardWeightedPairHistoryTwoKDepthTwo}{0.6105}
\newcommand{\NumRewardWeightedPairHistoryTwoKDepthFour}{0.6124}
\newcommand{\NumRewardWeightedPairHistoryTwoKDepthSix}{0.6134}
\newcommand{\NumRewardWeightedPairHistoryTwoKDepthEight}{0.6138}
\newcommand{\NumRewardWeightedPairHistoryEightKDepthOne}{0.6159}
\newcommand{\NumRewardWeightedPairHistoryEightKDepthTwo}{0.6189}
\newcommand{\NumRewardWeightedPairHistoryEightKDepthFour}{0.6207}
\newcommand{\NumRewardWeightedPairHistoryEightKDepthEight}{0.6219}
\newcommand{\NumRewardWeightedPairPretrained}{0.6215}

\newcommand{\NumInferenceBeamSize}{1024}
\newcommand{\NumInferenceMFU}{41\%}

\newcommand{\NumItemScorerNoneRecallThousand}{0.8656}
\newcommand{\NumItemScorerNoneTeacherRecallTen}{0.0381}
\newcommand{\NumItemScorerNoneTeacherRecallHundred}{0.1936}
\newcommand{\NumItemScorerNoneWeightedPair}{0.5478}
\newcommand{\NumItemScorerOneRecallThousand}{0.8615}
\newcommand{\NumItemScorerOneTeacherRecallTen}{0.5654}
\newcommand{\NumItemScorerOneTeacherRecallHundred}{0.7344}
\newcommand{\NumItemScorerOneWeightedPair}{0.5892}
\newcommand{\NumItemScorerFourRecallThousand}{0.8656}
\newcommand{\NumItemScorerFourTeacherRecallTen}{0.6005}
\newcommand{\NumItemScorerFourTeacherRecallHundred}{0.7501}
\newcommand{\NumItemScorerFourWeightedPair}{0.5937}
\newcommand{\NumDistillMAERecallThousand}{0.8615}
\newcommand{\NumDistillMAETeacherRecallTen}{0.5654}
\newcommand{\NumDistillMAETeacherRecallHundred}{0.7344}
\newcommand{\NumDistillMAEWeightedPair}{0.5892}
\newcommand{\NumDistillMSERecallThousand}{0.8671}
\newcommand{\NumDistillMSETeacherRecallTen}{0.5748}
\newcommand{\NumDistillMSETeacherRecallHundred}{0.7329}
\newcommand{\NumDistillMSEWeightedPair}{0.5915}
\newcommand{\NumDistillHuberRecallThousand}{0.8679}
\newcommand{\NumDistillHuberTeacherRecallTen}{0.5568}
\newcommand{\NumDistillHuberTeacherRecallHundred}{0.7220}
\newcommand{\NumDistillHuberWeightedPair}{0.5894}
\newcommand{\NumDistillKLRecallThousand}{0.8662}
\newcommand{\NumDistillKLTeacherRecallTen}{0.5452}
\newcommand{\NumDistillKLTeacherRecallHundred}{0.7151}
\newcommand{\NumDistillKLWeightedPair}{0.5860}
\newcommand{\NumRolloutBeamThirtyTwoRecallThousand}{0.8615}
\newcommand{\NumRolloutBeamThirtyTwoTeacherRecallTen}{0.5654}
\newcommand{\NumRolloutBeamThirtyTwoTeacherRecallHundred}{0.7344}
\newcommand{\NumRolloutBeamThirtyTwoWeightedPair}{0.5892}
\newcommand{\NumRolloutBeamSixtyFourRecallThousand}{0.8565}
\newcommand{\NumRolloutBeamSixtyFourTeacherRecallTen}{0.5661}
\newcommand{\NumRolloutBeamSixtyFourTeacherRecallHundred}{0.7359}
\newcommand{\NumRolloutBeamSixtyFourWeightedPair}{0.5901}
\newcommand{\NumRolloutBeamOneTwentyEightRecallThousand}{0.8616}
\newcommand{\NumRolloutBeamOneTwentyEightTeacherRecallTen}{0.5762}
\newcommand{\NumRolloutBeamOneTwentyEightTeacherRecallHundred}{0.7427}
\newcommand{\NumRolloutBeamOneTwentyEightWeightedPair}{0.5905}

\newcommand{\NumHistoryFullLongRecallThousand}{0.8722}
\newcommand{\NumHistoryFullLongTeacherRecallTen}{0.6586}
\newcommand{\NumHistoryFullLongTeacherRecallHundred}{0.7997}
\newcommand{\NumHistoryFullLongWeightedPair}{0.6029}
\newcommand{\NumHistoryCompressionRecallThousand}{0.8709}
\newcommand{\NumHistoryCompressionTeacherRecallTen}{0.6474}
\newcommand{\NumHistoryCompressionTeacherRecallHundred}{0.7893}
\newcommand{\NumHistoryCompressionWeightedPair}{0.6033}

\newcommand{\NumDistillBothRecallThousand}{0.8615}
\newcommand{\NumDistillBothTeacherRecallTen}{0.5654}
\newcommand{\NumDistillBothTeacherRecallHundred}{0.7344}
\newcommand{\NumDistillBothWeightedPair}{0.5892}
\newcommand{\NumDistillRolloutRecallThousand}{0.8610}
\newcommand{\NumDistillRolloutTeacherRecallTen}{0.5618}
\newcommand{\NumDistillRolloutTeacherRecallHundred}{0.7288}
\newcommand{\NumDistillRolloutWeightedPair}{0.5730}
\newcommand{\NumDistillImpressionRecallThousand}{0.8663}
\newcommand{\NumDistillImpressionTeacherRecallTen}{0.2983}
\newcommand{\NumDistillImpressionTeacherRecallHundred}{0.5281}
\newcommand{\NumDistillImpressionWeightedPair}{0.5872}

\newcommand{\NumOnlineLagMedian}{45 minutes}
\newcommand{\NumOnlineLagPNinetyNine}{60 minutes}
\newcommand{\NumModelSyncPeriod}{10 minutes}
\newcommand{\NumAttributionWindow}{15 minutes}

\newcommand{\NumProfileRealtimeWindow}{72 hours}
\newcommand{\NumProfileRealtimeCap}{1000}

\newcommand{\NumProdCandidateGenerators}{15}

\newcommand{\NumABVZeroActiveUsers}{0.56}
\newcommand{\NumABVZeroListeningTime}{1.81}
\newcommand{\NumABVZeroLikes}{0.00}
\newcommand{\NumABVOneActiveUsers}{0.73}
\newcommand{\NumABVOneListeningTime}{1.23}
\newcommand{\NumABVOneLikes}{2.71}
\newcommand{\NumABVTwoActiveUsers}{0.84}
\newcommand{\NumABVTwoListeningTime}{1.46}
\newcommand{\NumABVTwoLikes}{0.00}
\newcommand{\NumABArgusActiveUsers}{1.93}
\newcommand{\NumABArgusListeningTime}{3.69}
\newcommand{\NumABArgusLikes}{10.76}
\newcommand{\NumABSonaActiveUsers}{4.53}
\newcommand{\NumABSonaListeningTime}{6.30}
\newcommand{\NumABSonaLikes}{11.42}
\newcommand{\NumABSonaRepeats}{17.99}
\newcommand{\NumABSonaDeepUsers}{7.37}
\newcommand{\NumABSonaDays}{7}
\newcommand{\NumABSonaTrafficShare}{15\%}
\newcommand{\NumABExpLikeTrafficShare}{5\%}
\newcommand{\NumABPairwiseListeningTime}{2.05}
\newcommand{\NumABPairwiseLikes}{-1.03}
\newcommand{\NumABPairwiseRepeats}{7.43}
\newcommand{\NumABPairwiseActiveUsers}{1.51}
\newcommand{\NumABPairwiseDeepUsers}{2.61}
\newcommand{\NumABLikeOneListeningTime}{2.45}
\newcommand{\NumABLikeOneLikes}{6.56}
\newcommand{\NumABLikeOneRepeats}{12.89}
\newcommand{\NumABLikeOneActiveUsers}{1.83}
\newcommand{\NumABLikeOneDeepUsers}{2.87}
\newcommand{\NumABLikeTwoListeningTime}{2.28}
\newcommand{\NumABLikeTwoLikes}{11.93}
\newcommand{\NumABLikeTwoRepeats}{12.57}
\newcommand{\NumABLikeTwoActiveUsers}{1.53}
\newcommand{\NumABLikeTwoDeepUsers}{2.80}
\newcommand{\NumABExpSidTrafficShare}{4\%}
\newcommand{\NumABNoSidListeningTime}{2.60}
\newcommand{\NumABNoSidLikes}{11.13}
\newcommand{\NumABNoSidRepeats}{15.88}
\newcommand{\NumABNoSidActiveUsers}{1.98}
\newcommand{\NumABNoSidDeepUsers}{2.95}
\newcommand{\NumABSidListeningTime}{3.18}
\newcommand{\NumABSidLikes}{19.86}
\newcommand{\NumABSidRepeats}{20.52}
\newcommand{\NumABSidActiveUsers}{2.23}
\newcommand{\NumABSidDeepUsers}{3.80}
\newcommand{\NumABExpTeacherTrafficShare}{3\%}
\newcommand{\NumABTeacherListeningTime}{1.64}
\newcommand{\NumABTeacherLikes}{1.48}
\newcommand{\NumABTeacherRepeats}{4.68}
\newcommand{\NumABTeacherActiveUsers}{1.22}
\newcommand{\NumABTeacherDeepUsers}{1.83}
\newcommand{\NumABStackListeningTime}{3.63}
\newcommand{\NumABStackLikes}{-0.54}
\newcommand{\NumABStackRepeats}{5.02}
\newcommand{\NumABStackActiveUsers}{2.72}
\newcommand{\NumABStackDeepUsers}{4.06}
\newcommand{\NumABExpDistillTrafficShare}{8\%}
\newcommand{\NumABDistilledListeningTime}{1.62}
\newcommand{\NumABDistilledLikes}{7.12}
\newcommand{\NumABDistilledRepeats}{15.25}
\newcommand{\NumABDistilledActiveUsers}{1.41}
\newcommand{\NumABDistilledDeepUsers}{2.78}
\newcommand{\NumABDistilledTeacherListeningTime}{4.32}
\newcommand{\NumABDistilledTeacherLikes}{11.81}
\newcommand{\NumABDistilledTeacherRepeats}{14.93}
\newcommand{\NumABDistilledTeacherActiveUsers}{2.81}
\newcommand{\NumABDistilledTeacherDeepUsers}{5.11}

\newcommand{\NumABVsArgusRatio}{$2.35\times$}
\newcommand{\NumABListeningTime}{$+\NumABSonaListeningTime\%$}
\newcommand{\NumABActiveUsers}{$+\NumABSonaActiveUsers\%$}
\newcommand{\NumABLikes}{$+\NumABSonaLikes\%$}

\newcommand{\NumBaselineHistoryLength}{2k}
\newcommand{\NumBaselineEncoderLayers}{7}
\newcommand{\NumBaselineDecoderLayers}{2}
\newcommand{\NumBaselineRankingModuleLayers}{4}
\newcommand{\NumItemScorerOneLayerDepth}{1}
\newcommand{\NumBaselineSFTWeeks}{2}

\renewcommand{\headeright}{}
\renewcommand{\undertitle}{A Single-Model Generative Recommender for Yandex Music}

\title{Sona Technical Report}

\author{%
  Sona Team%
}
\date{}

\begin{document}
\maketitle

\vspace*{-0.30in}

\begin{abstract}
We introduce \Sona{}, a single-model generative recommender for Yandex Music.
In an online A/B test, \Sona{} replaced the entire production cascade --- more
than \NumProdCandidateGenerators{} candidate generators followed by pre-ranking
and ranking models that consume hundreds of features, including signals from
large transformer models such as Argus \citep{argus2025} and target-attention
scorers --- while significantly improving key engagement metrics.

The architecture of \Sona{} unifies candidate generation and ranking around a
shared user representation. Its encoder transforms the user's chronological
sequence of logged engagement events into hidden states consumed by both the
autoregressive decoder and the Ranking Module. The next-token-prediction and
distillation objectives jointly update the encoder, coupling generation and
ranking through the same user state. Neither \Sona{} nor its Teacher Ranker
uses hand-engineered features; both operate on logged event fields and learned
item representations. In the final \Sona{} configuration, the larger teacher
supplies ranking targets during training but is absent from serving, leaving
the encoder, decoder, and Ranking Module as a single deployed model.

We evaluate \Sona{} in an online A/B experiment using live traffic from My
Vibe on smart speakers, one of Yandex Music's largest recommendation surfaces.
On this surface, playback begins without the user specifying an artist, genre,
mood, or other preference. Relative to the production control, \Sona{} produced
statistically significant uplifts of \NumABActiveUsers{} in Active Users, the
primary metric, \NumABListeningTime{} in Total Listening Time, and
\NumABLikes{} in Likes. These effects were incremental to improvements retained
from preceding deployments. The Active Users uplift was
\NumABVsArgusRatio{} the increment previously delivered by Argus, the strongest
model deployed on this surface before \Sona{}. These results show that a single
jointly trained model can replace a mature multi-stage recommendation cascade
while improving recommendation quality on live traffic.
\end{abstract}

\definecolor{sonaYellow}{RGB}{255,204,0}
\definecolor{argusInk}{RGB}{78,78,75}
\definecolor{baselineVTwo}{RGB}{174,174,168}
\definecolor{baselineVOne}{RGB}{202,202,196}
\definecolor{baselineVZero}{RGB}{226,226,220}
\definecolor{chartRule}{RGB}{229,229,224}
\definecolor{chartText}{RGB}{57,57,54}

\pgfdeclarepatternformonly{large checker grid}
  {\pgfpointorigin}
  {\pgfqpoint{16mm}{16mm}}
  {\pgfqpoint{16mm}{16mm}}
  {%
    \pgfsetlinewidth{0.25pt}%
    \pgfpathrectangle{\pgfpointorigin}{\pgfqpoint{16mm}{16mm}}%
    \pgfpathmoveto{\pgfqpoint{8mm}{0mm}}%
    \pgfpathlineto{\pgfqpoint{8mm}{16mm}}%
    \pgfpathmoveto{\pgfqpoint{0mm}{8mm}}%
    \pgfpathlineto{\pgfqpoint{16mm}{8mm}}%
    \pgfusepath{stroke}%
  }

\newcommand{\gradientbar}[7][4.9pt]{%
  \draw[shade, top color=#2, bottom color=#2!12, draw=none]
    (axis cs:{#3-0.396},0)
    [rounded corners=1.6pt]
    -- (axis cs:{#3-0.396},{#4})
    -- (axis cs:{#3+0.396},{#4})
    [sharp corners]
    -- (axis cs:{#3+0.396},0)
    -- cycle;
  \node[
    font=\fontsize{5.3}{5.8}\selectfont,
    text=#6,
    anchor=north,
    inner sep=0pt,
    yshift=-#1
  ] at (axis cs:{#3},{#4}) {#5};
  \node[
    font=\fontsize{5.7}{6.3}\selectfont,
    text=chartText,
    anchor=south,
    yshift=1.6pt
  ] at (axis cs:{#3},{#4}) {#7};%
}

\newcommand{\gradientbarsmall}[5]{%
  \draw[shade, top color=#1, bottom color=#1!12, draw=none]
    (axis cs:{#2-0.396},0)
    [rounded corners=1.6pt]
    -- (axis cs:{#2-0.396},{#3})
    -- (axis cs:{#2+0.396},{#3})
    [sharp corners]
    -- (axis cs:{#2+0.396},0)
    -- cycle;
  \node[
    font=\fontsize{5.3}{5.8}\selectfont,
    text=chartText!60,
    anchor=south,
    yshift=1.2pt
  ] at (axis cs:{#2},{#3}) {#4};
  \node[
    font=\fontsize{5.7}{6.3}\selectfont,
    text=chartText,
    anchor=south,
    yshift=9.5pt
  ] at (axis cs:{#2},{#3}) {#5};%
}

\newcommand{\gradientzero}[3]{%
  \draw[chartText!35, line width=1pt, yshift=1.2pt]
    (axis cs:{#1-0.396},0) -- (axis cs:{#1+0.396},0);
  \node[
    font=\fontsize{5.3}{5.8}\selectfont,
    text=chartText!60,
    anchor=south,
    yshift=3.2pt
  ] at (axis cs:{#1},0) {#2};
  \node[
    font=\fontsize{5.7}{6.3}\selectfont,
    text=chartText,
    anchor=south,
    yshift=11.5pt
  ] at (axis cs:{#1},0) {#3};%
}

\begin{center}
  \vspace{1.4em}

  \makebox[\linewidth][c]{%
  \begin{tikzpicture}
    \begin{groupplot}[
      group style={
        group size=3 by 1,
        horizontal sep=0.65cm
      },
      width=0.3498\linewidth,
      height=6.0cm,
      ymin=0,
      xmin=0.45,
      xmax=5.55,
      xtick=\empty,
      ytick=\empty,
      tick style={draw=none},
      axis x line*=bottom,
      axis y line=none,
      axis line style={draw=chartText!55, line width=0.5pt},
      xlabel style={
        font=\scriptsize\bfseries,
        text=chartText,
        yshift=-4pt,
        align=center
      },
    ]
      \nextgroupplot[
        ymax=6.7,
        xlabel={Active Users\\[-1pt]
          {\fontsize{5.8}{6.3}\selectfont\color{sonaNavy}
           PRIMARY METRIC}}
      ]
        \addplot[draw=none, forget plot] coordinates {(1,0) (5,0)};
        \gradientbar[3.2pt]{baselineVZero}{1}{\NumABVZeroActiveUsers}{\NumABVZeroActiveUsers}{chartText!75}{V0}
        \gradientbar{baselineVOne}{2}{\NumABVOneActiveUsers}{\NumABVOneActiveUsers}{chartText!75}{V1}
        \gradientbar{baselineVTwo}{3}{\NumABVTwoActiveUsers}{\NumABVTwoActiveUsers}{chartText!85}{V2}
        \gradientbar{argusInk}{4}{\NumABArgusActiveUsers}{\NumABArgusActiveUsers}{white}{Argus}
        \gradientbar{sonaYellow}{5}{\NumABSonaActiveUsers}{\NumABSonaActiveUsers}{chartText}{\textbf{\Sona{}}}

      \nextgroupplot[
        ymax=8.1,
        xlabel={Total Listening Time\\[-1pt]
          {\fontsize{5.8}{6.3}\selectfont\color{white}PRIMARY METRIC}}
      ]
        \addplot[draw=none, forget plot] coordinates {(1,0) (5,0)};
        \gradientbar{baselineVZero}{1}{\NumABVZeroListeningTime}{\NumABVZeroListeningTime}{chartText!75}{V0}
        \gradientbar{baselineVOne}{2}{\NumABVOneListeningTime}{\NumABVOneListeningTime}{chartText!75}{V1}
        \gradientbar{baselineVTwo}{3}{\NumABVTwoListeningTime}{\NumABVTwoListeningTime}{chartText!85}{V2}
        \gradientbar{argusInk}{4}{\NumABArgusListeningTime}{\NumABArgusListeningTime}{white}{Argus}
        \gradientbar{sonaYellow}{5}{\NumABSonaListeningTime}{\NumABSonaListeningTime}{chartText}{\textbf{\Sona{}}}

      \nextgroupplot[
        ymax=13.0,
        xlabel={Likes\\[-1pt]
          {\fontsize{5.8}{6.3}\selectfont\color{white}PRIMARY METRIC}}
      ]
        \addplot[draw=none, forget plot] coordinates {(1,0) (5,0)};
        \gradientzero{1}{\NumABVZeroLikes}{V0}
        \gradientbar{baselineVOne}{2}{\NumABVOneLikes}{\NumABVOneLikes}{chartText!75}{V1}
        \gradientzero{3}{\NumABVTwoLikes}{V2}
        \gradientbar{argusInk}{4}{\NumABArgusLikes}{\NumABArgusLikes}{white}{Argus}
        \gradientbar{sonaYellow}{5}{\NumABSonaLikes}{\NumABSonaLikes}{chartText}{\textbf{\Sona{}}}
    \end{groupplot}

    \begin{scope}[on background layer]
      \fill[
        pattern=large checker grid,
        pattern color=chartText!7
      ] ([xshift=-2pt,yshift=-2pt]group c1r1.south west)
        rectangle
        ([xshift=2pt,yshift=2pt]group c3r1.north east);
    \end{scope}
  \end{tikzpicture}%
  }

  \vspace{0.6em}
  \parbox{0.84\linewidth}{\small
    \textbf{Figure 1 $\mid$ Successive transformer deployments in My Vibe on
    smart speakers.} V0, V1, V2, and Argus are successive
    generations of transformer-based recommendation models deployed on the
    surface before \Sona{}. Bars show incremental A/B uplift in percent;
    panels use independent scales. Each deployment adds its gain on top of
    those retained from all previous deployments. \Sona{} more than doubles
    Argus on the primary Active Users metric.}
\end{center}

\let\gradientbar\relax
\let\gradientbarsmall\relax
\let\gradientzero\relax

\newpage
\begingroup
\hypersetup{linkcolor=black}
\tableofcontents
\endgroup
\newpage

\section{Introduction}
\label{sec:intro}

We present \Sona{}, a generative recommender for Yandex Music that combines
candidate generation and ranking in a single model, evaluated in
online A/B experiments on live traffic. Its served
architecture is a transformer over the user's chronological event
history with two output modules: a \emph{decoder} and a \emph{Ranking
Module}. In the final experiment this single model replaces the candidate
generation, pre-ranking, and ranking stages of a mature recommendation
cascade of separately trained models. The report describes the
architecture, training recipe, and infrastructure that allowed this
model to surpass the cascade on live traffic.

Yandex Music is a subscription music-streaming service built around
personalized listening: alongside search, curated playlists, and
editorial content, the home feed, radio-style stations, and the
personal My Vibe stream are all driven by recommendations, and
recommended playback accounts for a large share of total listening
time. The service runs on mobile and desktop applications, the web player,
TVs, and smart speakers with a voice assistant. My Vibe on smart speakers is
our test platform for \Sona{}. Playback begins without the user specifying an
artist, genre, mood, or other preference. This creates a pure-recommendation
setting for measuring recommendation quality; the surface is also one of the
largest by listening time.

Several properties distinguish the music domain from short-video and
e-commerce recommendation and shape the design of \Sona{}.
Consumption is largely passive: the stream plays continuously while
the user attends to something else, so feedback arrives as plays,
skips, and occasional likes rather than deliberate choices. Repeated
listens are the norm rather than the exception --- returning to a
familiar track is often exactly what the user wants, so repetition
is signal to be modeled, not redundancy to be filtered out. These
two properties combine into the central tension of the domain:
familiar tracks reliably sustain time spent, while discovery of new
favorites pays off over a longer horizon, so the recommender must
continuously balance time spent against novelty.

\Sona{} formulates next-track recommendation as conditional sequence
generation. The encoder consumes typed engagement events --- including
plays, skips, likes, and dislikes --- together with their track
identities and request context. Following the Semantic ID formulation
of \citet{rajput2023} and its production successors
\citep{hstu2024, onerec2025, oxygenrec2025, gpr2025}, the decoder
emits each recommendation as a short tuple of discrete Semantic IDs (SIDs)
(\Cref{sec:arch:tokenizer}). Each tuple is expanded into the catalog
tracks that share it, and the Ranking Module ranks the resulting candidates
(\Cref{sec:arch:decoder}). The two modules share a user encoder that is
evaluated once per request, allowing generation and ranking to run in
a single served model. Across training and serving, inputs are
constructed from logged event fields and learned Semantic IDs; no
component uses hand-crafted features.

The Ranking Module's training supervision comes from a Teacher Ranker
(\Cref{sec:teacher:arch}), a sequential transformer over the same event
vocabulary as the encoder with a cross-attention candidate scorer.
We train it with a two-stage initial recipe: next-item-prediction pre-training
learns the sequence representation, and multi-head ranking fine-tuning
adapts it to engagement objectives; the trained model is the
distillation teacher. Few production reports describe
this pre-training --- Argus \citep{argus2025} and GPSD
\citep{gpsd2025} among them --- and our ablations make it the
load-bearing stage: it is what lets the teacher alone, one transformer
with no hand-engineered features, replace a ranking model driven by
hundreds of features, among them large-scale transformer models such
as Argus \citep{argus2025} and target-attention scorers
(\Cref{sec:eval:online}).

The served model is connected to a continuous online-training loop:
an emitted listening event is aggregated into its session over a
window of \NumAttributionWindow{}, flows through real-time processing
into a queue consumed by the training process, and fresh model weights
are delivered to production every \NumModelSyncPeriod{}. End to end,
the latency from event emission to an updated production model is
\NumOnlineLagMedian{} at the median and \NumOnlineLagPNinetyNine{} at
the 99th percentile (\Cref{sec:infra:online}).

We validate \Sona{} both offline and online. For offline validation, we use
\emph{Teacher Recall@$k$} to measure how closely the Ranking Module
preserves the teacher's ordering over the same generated candidates
(\Cref{sec:eval:protocol,sec:eval:unified}). We then evaluate intermediate and
final serving configurations in five online experiments presented in
conceptual order. In the final experiment in My Vibe on smart speakers,
\Sona{} increases
Active Users by \NumABActiveUsers{}, Total Listening Time by
\NumABListeningTime{}, and Likes by \NumABLikes{}
(\Cref{sec:eval:online}). On the primary Active Users metric, this uplift is
\NumABVsArgusRatio{} the increment previously delivered by Argus
\citep{argus2025}, the strongest model previously deployed there.

\section{\texorpdfstring{\Sona{}}{Sona} Overview}
\label{sec:overview}

\subsection{Design Philosophy}
\label{sec:philosophy}

A small number of ideas guided us throughout the work on \Sona{} ---
in choosing the formulation, weighing alternatives, and deciding what
to ship. We state each idea and the reasoning behind it below; the
rest of the report describes the architecture, training recipe, and
infrastructure they shaped.

\paragraph{One model for generation and ranking.}
Large language models demonstrated that a single model trained
end-to-end on a large corpus can absorb functions previously split
across specialized components, and single-model generative recommenders
carried this recipe into production recommendation
\citep{onerec2025, oxygenrec2025, gpr2025}. Motivated by these
results, we set out to explore the potential of a single-model
recommender in the music domain.

\paragraph{One user representation, shared across tasks.}
We are guided by the idea that a single user representation should
serve every recommendation task built on top of it: the encoder cost
is paid once per request, and the resulting representation is reused
for both generation and ranking. Recent industry deployments show
that candidate generation and item-level ranking can share one
backbone at production scale \citep{unipinrec2026, recochain2026}.
We therefore adopt Gryphon \citep{gryphon2026}, a unified
generation-and-ranking design, as the architecture of \Sona{}
(\cref{sec:arch:musicgryphon}).

\paragraph{No hand-engineered features.}
Across machine-learning domains, models given scale and data have
matched pipelines that encode domain structure by hand: image
transformers reach the accuracy of architectures built around
convolutional inductive bias once pre-training data is plentiful
\citep{vit2021}, and language-model quality follows smooth scaling
curves in parameters and data \citep{kaplan2020}. Recommendation
reproduced the pattern: sequence models over raw engagement events
scale the same way \citep{argus2025}, and transformer rankers over
those sequences reach the quality of feature-rich production rankers
\citep{hstu2024, ligr2025}. We
therefore build \Sona{} and its teacher on logged event fields and
learned Semantic IDs alone, and set out to explore how far that
reaches in music recommendation.

\paragraph{Dense supervision for ranking without hand-engineered features.}
Next-token prediction and action prediction supply supervision at
very different densities: next-token prediction produces a training
target at every position of every sequence, while explicit feedback
events are sparse. A strong ranking model without hand-engineered
features must recover from data the structure such features would
otherwise encode, and this lack of built-in inductive bias calls for
training on large-scale datasets. Autoregressive training over
engagement sequences is widely adopted as the way to leverage
datasets of this scale \citep{hstu2024, mtgr2025}. We follow this
recipe, and we initially train the Teacher Ranker in two stages ---
next-item-prediction pre-training followed by multi-head ranking
fine-tuning \citep{argus2025}; our ablations find the pre-training
stage crucial for the teacher's quality (\Cref{tab:eval:reward}).

\paragraph{Compress a year of history into a teacher, distill it
densely into the student.}
For encoder--decoder models trained with next-token prediction on
chronologically organized data, each training sample carries the
user's full interaction history, so both the dataset's size on disk
and the compute spent encoding it grow far faster than the number of
training samples: every sample re-encodes the history from scratch to
supervise a handful of new targets. This cost makes such datasets
unsuitable for a multi-month span of dates. We
therefore train a large-scale ranking model autoregressively on a
dataset spanning a year of engagement logs, compressing that history
into its weights, and distill its knowledge densely into the Ranking
Module, which is trained jointly with the next-token-prediction task
on a dataset covering a much shorter range of dates
(\cref{sec:teacher:arch,sec:train:rescorer}).

\paragraph{Aligning generation and ranking through Rollout Distillation.}
We use beam-search rollouts from the current decoder as the candidate source
for Rollout Distillation (\Cref{sec:train:rescorer}). The frozen teacher scores
the generated candidates, and the Ranking Module learns to reproduce those
scores. Because the rollouts come from the current decoder distribution, the
distillation signal follows the candidate distribution the model produces
at that point in training. The decoder and Ranking Module also share the
user encoder, so this signal shapes the representation that conditions both
tasks. In this way, Rollout Distillation aligns generation and ranking inside
the joint training objective.

\newpage
\subsection{Overall Framework}
\label{sec:framework}

\begin{figure}[H]
  \centering
  \includegraphics[width=\linewidth]{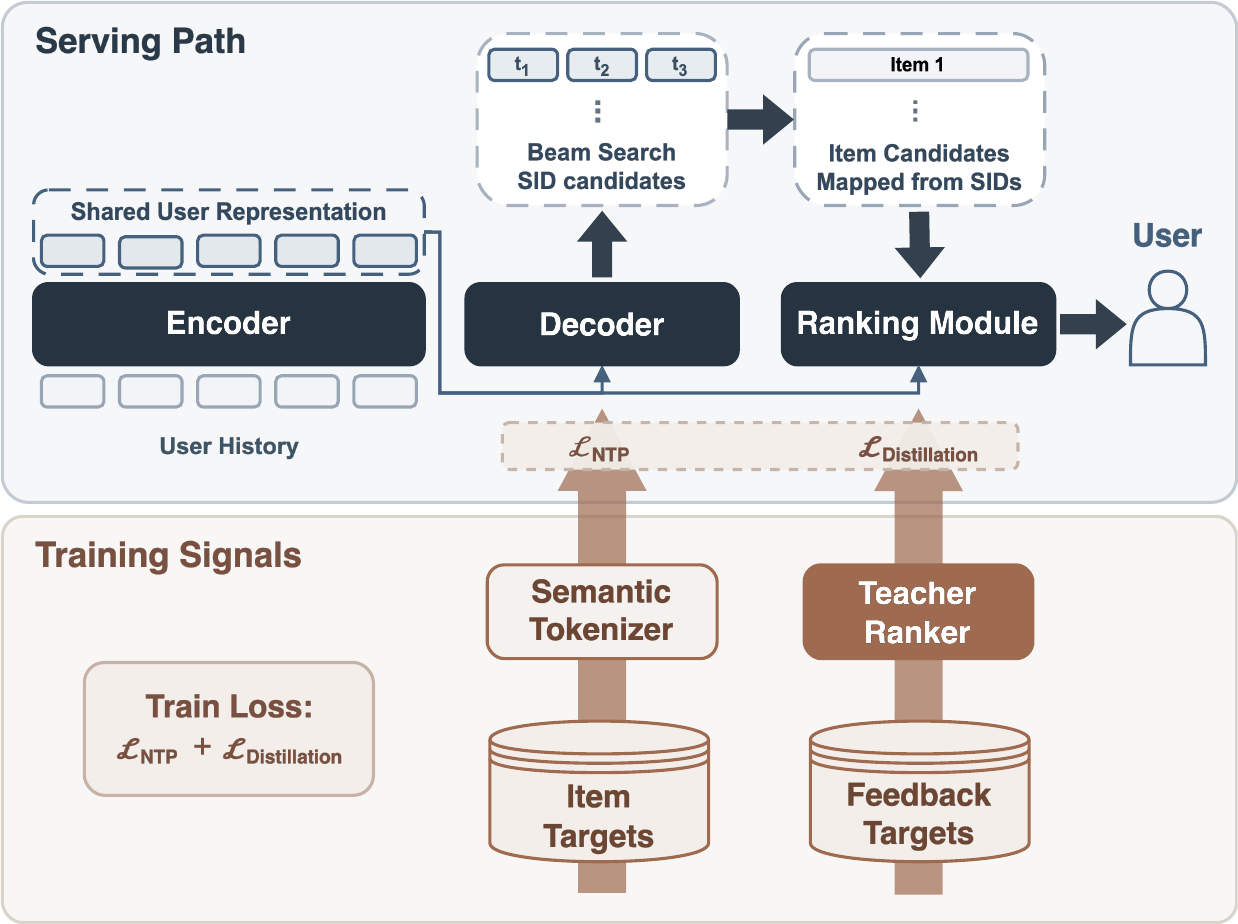}
  \caption{\Sona{} architecture and data flow.}
  \label{fig:sona-overview}
\end{figure}

\Cref{fig:sona-overview} summarizes \Sona{}'s serving path and training
signals. At serving time, the encoder processes the user's chronological
event history once, producing a representation shared by the decoder and
Ranking Module. Following the generative-recommendation paradigm, the decoder
autoregressively produces a beam of short Semantic ID tuples
\citep{rajput2023, hstu2024, onerec2025, oxygenrec2025, gpr2025}. A catalog
index maps each tuple to one or more tracks, and the Ranking Module uses the
shared encoder states to score and order the resulting candidates.

During training, the semantic tokenizer converts each target item into its
Semantic ID sequence, providing the next-token-prediction objective for the
decoder. The frozen teacher ranker scores decoder-generated candidates and
provides distillation targets for the Ranking Module. Both losses
back-propagate through the shared encoder, so its representation learns from
generation and ranking. In the final \Sona{} configuration, the tokenizer and
teacher are used only for training; serving consists of the encoder, decoder,
deterministic SID-to-item mapping, and Ranking Module. The teacher was
temporarily added to limited-traffic treatments in Online Experiments~1--4
(\Cref{sec:eval:online}).

\Cref{tab:architecture-components} summarizes the role and lifecycle of each
component.

\section{\texorpdfstring{\Sona{}}{Sona} Architecture}
\label{sec:architecture}

This section details the components of \Sona{}: the semantic tokenizer,
user-event representation, shared encoder with History Compression,
autoregressive decoder, and Ranking Module.

\begin{table}[H]
  \renewcommand{\thetable}{2.1}
  \centering
  \footnotesize
  \renewcommand{\arraystretch}{1.08}
  \begin{tabularx}{\textwidth}{@{}
    >{\raggedright\arraybackslash}p{0.17\textwidth}
    >{\raggedright\arraybackslash}p{0.21\textwidth}
    >{\raggedright\arraybackslash}X
    >{\raggedright\arraybackslash}p{0.25\textwidth}@{}}
    \toprule
    \rowcolor{sonaNavy}
    \textcolor{white}{\textbf{Component}}
      & \textcolor{white}{\textbf{Consumes}}
      & \textcolor{white}{\textbf{Produces / role}}
      & \textcolor{white}{\textbf{Optimization and lifecycle}} \\
    \midrule
    \arrayrulecolor{white}
    \rowcolor{sonaTrainingTint}
    \textcolor{sonaBrown}{\textbf{Semantic Tokenizer}}
      & Dense item representations
      & A fixed SID tuple for each item, defining decoder targets and the
        SID-to-item index
      & Offline; fixed during training and serving \\
    \specialrule{0.8pt}{0pt}{0pt}

    \rowcolor{sonaServingTint}
    \textcolor{sonaNavy}{\textbf{Shared User Encoder}}
      & Chronological user events
      & Hidden states shared by generation and ranking
      & Training and serving; receives gradients from both next-token and
        distillation objectives \\
    \specialrule{0.8pt}{0pt}{0pt}

    \rowcolor{sonaServingTint}
    \textcolor{sonaNavy}{\textbf{Semantic ID Decoder}}
      & Encoder states and preceding SID tokens
      & Beam of candidate SID tuples
      & Next-token training; served \\
    \specialrule{0.8pt}{0pt}{0pt}

    \rowcolor{sonaServingTint}
    \textcolor{sonaNavy}{\textbf{SID-to-item Mapper}}
      & Generated SIDs and catalog index
      & Catalog candidates
      & Deterministic; no learned parameters \\
    \specialrule{0.8pt}{0pt}{0pt}

    \rowcolor{sonaServingTint}
    \textcolor{sonaNavy}{\textbf{Ranking Module}}
      & Encoder states and candidate item / SID features
      & Multi-feedback scores and final ordering
      & Teacher-distilled; served \\
    \specialrule{0.8pt}{0pt}{0pt}

    \rowcolor{sonaTrainingTint}
    \textcolor{sonaBrown}{\textbf{Teacher Ranker}}
      & Event histories and candidate items
      & Item-level distillation targets
      & Separately trained and frozen \\
    \arrayrulecolor{sonaNavy}
    \bottomrule
    \arrayrulecolor{black}
  \end{tabularx}
  \caption{Final \Sona{} serving components and training-only supervision.}
  \label{tab:architecture-components}
\end{table}

\subsection{Semantic Tokenizer}
\label{sec:arch:tokenizer}

Generative recommenders face large and continuously growing item
spaces, which makes the direct generation of atomic item identifiers
computationally and architecturally infeasible. Items are therefore
tokenized into short coarse-to-fine sequences of discrete codes ---
\emph{Semantic IDs} --- drawn from a small fixed vocabulary
\citep{rajput2023}. This keeps the output space compact and stable as
the catalog grows, and items with similar content or behavioral
patterns share leading codes, which transfers knowledge across related
tracks and improves generalization to newly added items
\citep{onerec2025}.

\begin{figure}[t!]
      \centering
      \includegraphics[width=\linewidth]{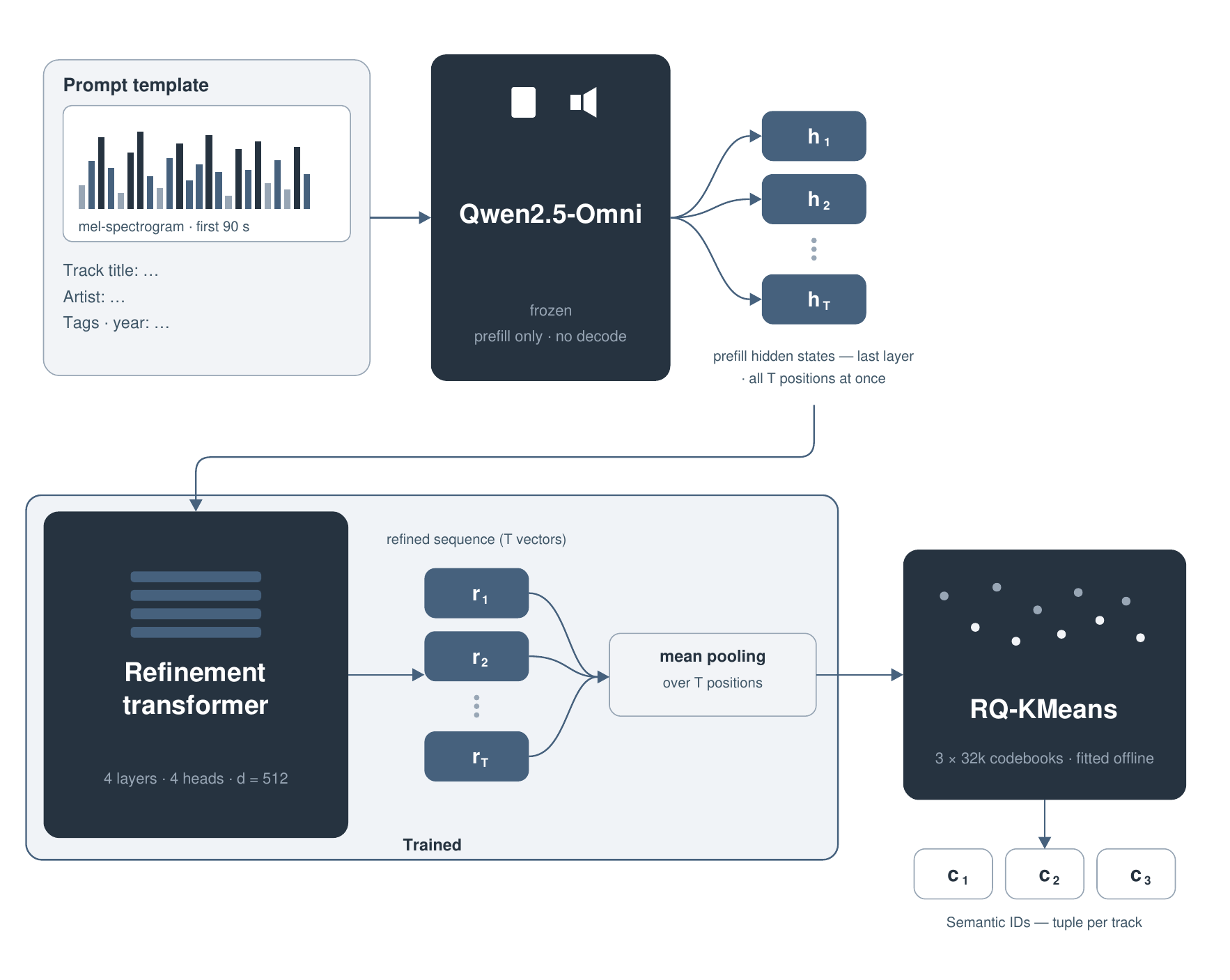}
      \caption{Tokenization pipeline. The frozen Qwen2.5-Omni runs in
      prefill-only mode --- no decoding.}
      \label{fig:sona-tokenization}
\end{figure}

The tokenizer realizes this scheme for the music catalog: it assigns
every track a fixed tuple of \NumSemIDLevels{} Semantic IDs, drawn from
\NumSemIDLevels{} codebooks of \NumCodebookSize{} entries each. These
tuples are the decoder's output vocabulary and define the SID-to-item
index used at serving. The tokenizer is built offline in three stages
(\Cref{fig:sona-tokenization}): content features from the hidden states
of a frozen multimodal LLM, a refinement transformer trained on
collaborative pairs, and a residual quantizer.

For each eligible track, the frozen Qwen2.5-Omni \citep{qwen25omni}
encodes the mel-spectrogram of the first \(90\) seconds of audio
together with a textual prompt carrying the title, artists, tags, and
additional metadata. Its last-layer states --- the track's
\emph{content features} --- are refined by a four-layer transformer
with four attention heads and model dimension \(512\), following
QARM-style pipelines \citep{rqkmeans2024}, and mean-pooled into a
single \NumTokenizerEmbDim{}-dimensional \emph{item embedding} \(z\)
for the track. Once the transformer is trained and frozen, \(z\) is
the only representation the quantizer receives.

\paragraph{Collaborative refinement.} Content features alone leave
substitutability implicit: items that listeners treat as
interchangeable are not necessarily neighbors in the content embedding
space \citep{letter2024,plum2025}. The refinement transformer is
therefore trained on collaboratively related track pairs, so that
tracks with similar listening patterns receive nearby item embeddings
--- and, after quantization, shared code prefixes. Its loss combines
two terms,
\[
    \mathcal{L}
    =
    \mathcal{L}_{\mathrm{InfoNCE}}
    +
    \lambda_{\mathrm{align}}
    \mathcal{L}_{\mathrm{align}} .
\]
A batch of \(m\) collaborative pairs yields \(2m\) refined item embeddings,
written \(z_a\) for track \(a\) of the batch. Each serves once as the
anchor; its paired track \(z_{a^{+}}\) is the positive and the
remaining \(2m-2\) embeddings are negatives:
\[
  \mathcal{L}_{\mathrm{InfoNCE}}
  =
  -\frac{1}{2m}
  \sum_{a=1}^{2m}
  \log
  \frac{
    \exp\bigl(\mathrm{sim}(z_a, z_{a^{+}}) / \tau\bigr)
  }{
    \displaystyle\sum_{b \neq a}
    \exp\bigl(\mathrm{sim}(z_a, z_b) / \tau\bigr)
  },
\]
with cosine similarity \(\mathrm{sim}\) and temperature \(\tau\). The
alignment term keeps each item embedding close to \(\bar h_a\), the
mean of the same track's content features,
\[
  \mathcal{L}_{\mathrm{align}}
  =
  \frac{1}{2m}
  \sum_{a=1}^{2m}
  \bigl\lVert z_a - \bar h_a \bigr\rVert^2,
\]
with weight \(\lambda_{\mathrm{align}} = 0.1\).

\paragraph{Collaborative pairs.} Pairs are mined from two sources. The
\emph{target-wise} stream collects co-served item sets from the
production recommender --- groups of items presented together in one
recommendation response. The production recommender does not place
tracks by the same artist in a single response, so these co-occurrences
are cross-artist by construction. The \emph{NPMI} stream scores
co-listening of two tracks \(i\) and \(j\) with normalized pointwise
mutual information,
\[
  \mathrm{NPMI}(i,j)
  = \frac{1}{-\log p(i,j)}\,
    \log \frac{p(i,j)}{p(i)\,p(j)} ,
\]
where \(p(i,j)\) is the probability that \(i\) and \(j\) co-occur in a
user's listening log over a three-month window and \(p(i)\), \(p(j)\)
are the marginals. Only high-NPMI pairs in which both tracks belong to
the same artist are kept, so the two streams are disjoint.
\Cref{tab:arch:collab-pairs} summarizes the sources; the streams are
concatenated and shuffled before training.

\begin{table}[H]
  \centering
  \small
  \begin{tabular}{lcc}
    \toprule
    \textbf{Pair source} & \textbf{Mining window} & \textbf{Pairs} \\
    \midrule
    Target-wise (co-served, cross-artist) & 3 weeks  & \(220\)--\(240\)M \\
    NPMI (co-listening, same-artist)      & 3 months & \(20\)M \\
    \bottomrule
  \end{tabular}
  \caption{Collaborative pair sources for the tokenizer InfoNCE objective.}
  \label{tab:arch:collab-pairs}
\end{table}

\paragraph{Residual quantization.} The refined item embedding \(z\) is mapped
to its Semantic ID tuple by residual K-means:
\NumSemIDLevels{} levels of \NumCodebookSize{} centroids, initialized
with k-means\(++\) and assigned by Euclidean distance. Each level
quantizes the residual of the previous one, so the tuple is a
coarse-to-fine path through the codebooks. The comparisons that
selected this configuration --- the codebook size and the item
representation supplied to the quantizer --- are reported in
\Cref{sec:eval:tokenizer}.

\subsection{User History Representation}
\label{sec:arch:history}

The encoder input is built from the user's logged engagement history.
For a request, the history is the chronological sequence
\[
  u = (e_1, \ldots, e_T), \qquad T \leq T_{\max},
\]
where each \(e_t\) is one past engagement event. The same selection rule
is applied in training and serving. Histories shorter than \(T_{\max}\)
are padded with a single learned embedding shared across the missing
positions.

\paragraph{Event embedders.} Each event \(e_t\) is embedded by three
embedders, one per feature group: an \emph{item embedder} over item
identity, a \emph{context embedder} over the request surface, and a
\emph{feedback embedder} over the observed response. \Cref{tab:arch:backbone:features}
lists the source signals each embedder reads and the transformations that
produce its inputs. Every input is a logged attribute; no
hand-engineered features are used.

\begin{table}[H]
  \centering
  \small
  \begin{tabularx}{0.98\linewidth}{lXX}
    \toprule
    \textbf{Embedder} & \textbf{Source signals} & \textbf{Construction} \\
    \midrule
    Item
      & track ID, artist ID, track duration
      & hash track ID three times and artist ID twice into one shared
        embedding table; bucket and embed track duration \\
    \addlinespace[0.5em]
    Context
      & smart-speaker flag, organic-feed flag
      & embed both request-surface flags \\
    \addlinespace[0.5em]
    Feedback
      & like flag, played time
      & embed the like flag; divide played time by track duration into
        the completion ratio, then bucket and embed it \\
    \bottomrule
  \end{tabularx}
  \caption{Historical-event features and their transformations before
  the three input embedders.}
  \label{tab:arch:backbone:features}
\end{table}

\paragraph{Continuous features.} Two continuous quantities reach the
embedders, and both are bucketed and embedded rather than used as raw
scalars: track duration, and the completion ratio
\[
  \rho_t = \frac{p_t}{\max(d_t, \epsilon)},
\]
where \(p_t\) is played time and \(d_t\) track duration.

\paragraph{Event token.} The item embedder concatenates its five hashed
identity features with the track-duration embedding and projects the
result to \(\mathbb{R}^{d}\); the context and feedback embedders embed each
of their attributes directly at width \(d\) and sum them. The event
token is the sum of the three embedder outputs:
\[
  x_t =
    \operatorname{ItemEmb}(e_t)
    + \operatorname{ContextEmb}(e_t)
    + \operatorname{FeedbackEmb}(e_t).
\]

\paragraph{Encoder input.} A learned \textsc{cls} token is prepended as
a global summary position, giving
\[
  X =
  \left[
    x_{\mathrm{CLS}},\,
    x_1,\,
    \ldots,\,
    x_T
  \right].
\]

\subsection{Encoder}
\label{sec:arch:encoder}

For a full encoder, let \(H\) denote the hidden-state matrix computed
from the event sequence \(X\) of \Cref{sec:arch:history},
\[
  H = \mathrm{Enc}_{\theta}(X), \qquad
  H \in \mathbb{R}^{(T+1) \times d}.
\]
The decoder (\Cref{sec:arch:decoder}) and Ranking Module
(\Cref{sec:arch:musicgryphon}) consume a shared encoder memory \(K\),
computed once per request. The symbols distinguish the encoder's direct
output from the memory exposed to downstream modules: \(K=H\) for a full
encoder, whereas History Compression does not construct a full-depth \(H\)
over all positions and instead forms \(K=[X_O;H_R]\). Thus \(K\) is the
common downstream interface in both cases.

\begin{figure}[t]
  \centering
  \includegraphics[width=\linewidth]{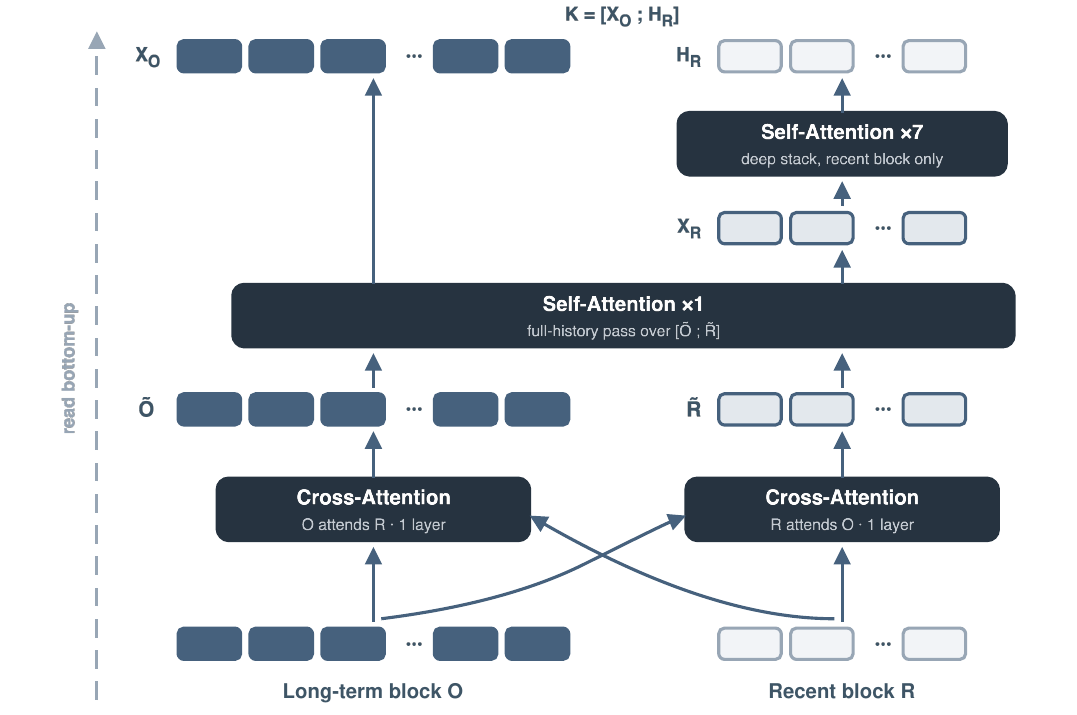}
  \caption{History Compression, read bottom-up. Dark: the long-term block
  \(O\) (\(n_{\mathrm{o}}\) events); light: the recent block \(R\)
  (\(n_{\mathrm{r}}\) events).}
  \label{fig:sona-histcompress}
\end{figure}

\paragraph{History Compression.} Attention cost grows quadratically
with history length, and keeping long histories affordable under
serving-latency budgets is a recurring constraint for industrial user
models \citep{sum2024, llatte2026}. \Sona{} therefore spends its depth
unevenly. The history of up to \(N\) events is split into a
\emph{recent} block \(R\) of the \(n_{\mathrm{r}}\) most recent events
and a \emph{long-term} block \(O\) of the preceding
\(n_{\mathrm{o}} = N - n_{\mathrm{r}}\) events, and the encoder
applies four stages (\Cref{fig:sona-histcompress}). History
Compression is used only in \Sona{}; the Teacher Ranker
(\Cref{sec:teacher:arch}) encodes its history in full.

\begin{enumerate}
  \item \textbf{Cross-attention.} Two single-layer cross-attention
        blocks let each block read the other --- \(O\) attends to
        \(R\) and \(R\) attends to \(O\) --- giving enriched blocks
        \(\tilde O\) and \(\tilde R\).
  \item \textbf{Full-history pass.} One self-attention layer runs
        over the concatenation \([\tilde O;\tilde R]\), producing block
        representations \(X_O\) and \(X_R\). This is the only stage
        that mixes information across the whole history.
  \item \textbf{Deep stack, recent block only.} A stack of
        \NumBackboneLayers{} self-attention layers, which carries most
        of the encoder's capacity, refines \(X_R\) over the
        \(n_{\mathrm{r}}\) recent events only, yielding \(H_R\).
  \item \textbf{Concatenated read-out.} Modules that cross-attend to
        the encoder read
        \[
          K = \bigl[\,X_O \,;\, H_R\,\bigr],
        \]
        the shared encoder memory: all \(N\) positions are visible
        downstream, while only the recent \(n_{\mathrm{r}}\) carry the
        deep encoding.
\end{enumerate}

\paragraph{Cost.} A full \(L\)-layer encoder over \(N\) events spends
\(\Theta(L\,N^2 d)\) in attention; ours pays the deep cost only over
the recent block, reducing the attention cost to
\(\Theta\bigl(
  (L\,n_{\mathrm{r}}^2 + N^2 + 2\,n_{\mathrm{o}} n_{\mathrm{r}})\,d
\bigr)\). The
saving comes from where capacity is spent, not from discarding
history. We adopt \(N=\NumCompressMaxLen{}\) and
\(n_{\mathrm{r}}=\NumCompressRecentLen{}\); this preserves most of the
downstream performance of a full transformer over the same history at
about half its inference cost (\Cref{sec:eval:unified}).

\paragraph{Block design.} All transformer blocks use RMSNorm
\citep{zhang2019rmsnorm} in a pre-normalization layout
\citep{xiong2020layernorm}, SwiGLU feed-forward blocks
\citep{shazeer2020}, and RoPE positional embeddings \citep{rope2021};
the complete model cards are given in
\Cref{tab:hparams:encoder,tab:hparams:decoder}.

\subsection{Decoder}
\label{sec:arch:decoder}

The tokenizer (\Cref{sec:arch:tokenizer}) maps each
catalog track \(v\) to an \(L\)-level Semantic ID tuple
\[
  s(v) = (s_1(v), \ldots, s_L(v)), \qquad
  L = \NumSemIDLevels{}.
\]
The decoder is a shallow autoregressive transformer over the codebook
vocabulary (\Cref{tab:hparams:decoder}). It starts from \textsc{bos} and generates
the tuple one level at a time, conditioned on the encoded history:
\[
  p_{\theta}(s_1,\ldots,s_L \mid u)
  =
  \prod_{\ell=1}^{L}
    p_{\theta}\!\left(s_\ell \mid s_{<\ell}, K\right).
\]

\paragraph{From tuples to tracks.} At serving time, constrained beam
search masks invalid prefixes with the tokenizer trie, so each completed
tuple maps to at least one catalog track. The codebook is not
collision-free: a typical SID tuple is shared by a small number of
catalog tracks, as is intrinsic to the Semantic ID formulation of
\citet{rajput2023}. Each completed tuple is therefore expanded into the
full collision class of tracks that quantize to it, and the Ranking
Module orders the expanded candidates
(\Cref{sec:arch:musicgryphon}).

\subsection{Ranking Module}
\label{sec:arch:musicgryphon}

\begin{figure}[t]
  \centering
  \includegraphics[width=\linewidth]{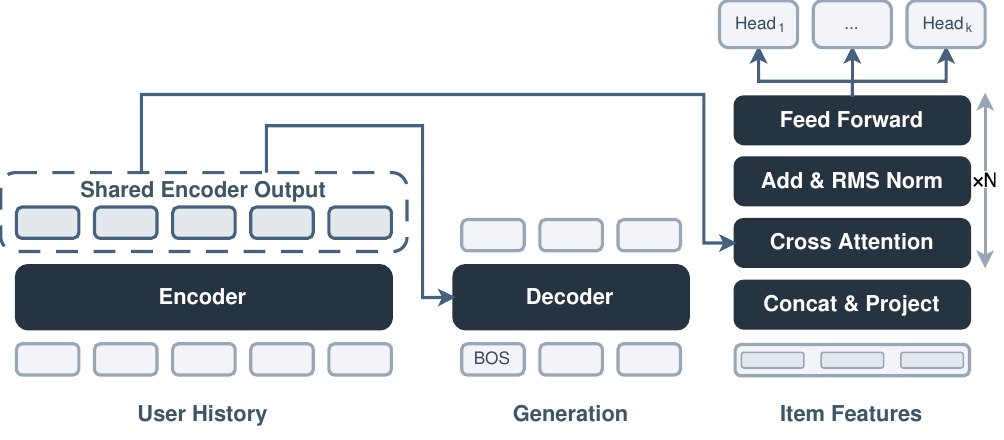}
  \caption{Unified generation-and-ranking architecture: the decoder
  and the Ranking Module attend to the same encoded user history.}
  \label{fig:sona-musicgryphon}
\end{figure}

The Ranking Module scores each candidate that beam search proposes
against the same encoder memory \(K\) the decoder attends to, following
the unified generation-and-ranking design of Gryphon
\citep{gryphon2026}: Semantic ID generation proposes candidates with
high recall, and the Ranking Module orders them
(\Cref{fig:sona-musicgryphon}). The history is encoded once, so scoring
adds no second encoder pass.

\paragraph{Candidate representation.} Each candidate is represented by
three kinds of features: its item ID, its Semantic ID codes
\(c_1, \ldots, c_L\) taken independently, and their prefix
n-gram features \citep{zheng2025enhancing}. Following the
unified-embedding scheme of \citet{unifiedemb2023}, every feature is
mapped by several independent hash functions into the embedding table
shared with the encoder
(\Cref{sec:arch:history}). The looked-up rows are concatenated and
projected into a single dense vector \(e_i\).

\paragraph{Scoring.} The Ranking Module largely follows the decoder
architecture, with the candidate embedding \(e_i\) as its input. A stack of
pre-norm cross-attention blocks refines this embedding against the encoded
history \citep{prenorm2020},
\[
  \tilde{e}_{i}
  =
  \mathrm{CrossAttnBlocks}(e_i, K).
\]
A feed-forward scoring layer maps the result to \(n\) head scores,
\[
  \mathbf{s}_{i}
  =
  \mathrm{Head}(\tilde{e}_{i}) \in \mathbb{R}^{n},
\]
one per Teacher Ranker head, trained by distillation from the teacher
jointly with the decoder's next-token objective
(\Cref{sec:train:rescorer}).

\paragraph{Combined score.} A fixed weighted combiner collapses the
\(n\) head scores into a single scalar, and the expanded candidates are
ordered by it to produce the final recommendation list.

\section{\texorpdfstring{\Sona{}}{Sona} Training}
\label{sec:training}

\subsection{Dataset Design}
\label{sec:train:data}

The dataset is request-level: its unit is a single served
recommendation request. A sample carries the user's chronological event
history up to the request and the \emph{impressions} --- the items
served in response, stamped with the request timestamp and the feedback
they subsequently received. History events carry exactly the logged
attributes of \Cref{tab:arch:backbone:features}. Samples with an empty
history are dropped, and the sample stream inherits the traffic gates
of the online loop (\Cref{sec:infra:online}).

\paragraph{Supervision.} All positive items of a request supervise the
model against the shared user history: the tokenizer maps each
impression to its Semantic ID tuple, and the positives form the packed
target set \(C^{+}(u)\) of the next-token objective --- several targets
per request, one encoder pass (\Cref{sec:train:pretrain}). The full
impression list is the impression candidate source of the
distillation objective, and beam-search rollouts supply its second
source. Impressions absent from the tokenizer's catalog snapshot are
masked out of the loss.

\paragraph{Time ordering.} The dataset is organized chronologically.
For initial training, target requests are split into day-level tables
and the trainer consumes them from oldest to newest. Once the model
enters online training (\Cref{sec:infra:online}), it consumes the
request stream directly in event order, with no splitting into
timestamp chunks. In both stages training sweeps time forward, and
every sample conditions only on events that precede its request.

\subsection{Training Framework}
\label{sec:train:pretrain}
\label{sec:train:rescorer}

\begin{figure}[t!]
  \centering
  \includegraphics[width=\linewidth]{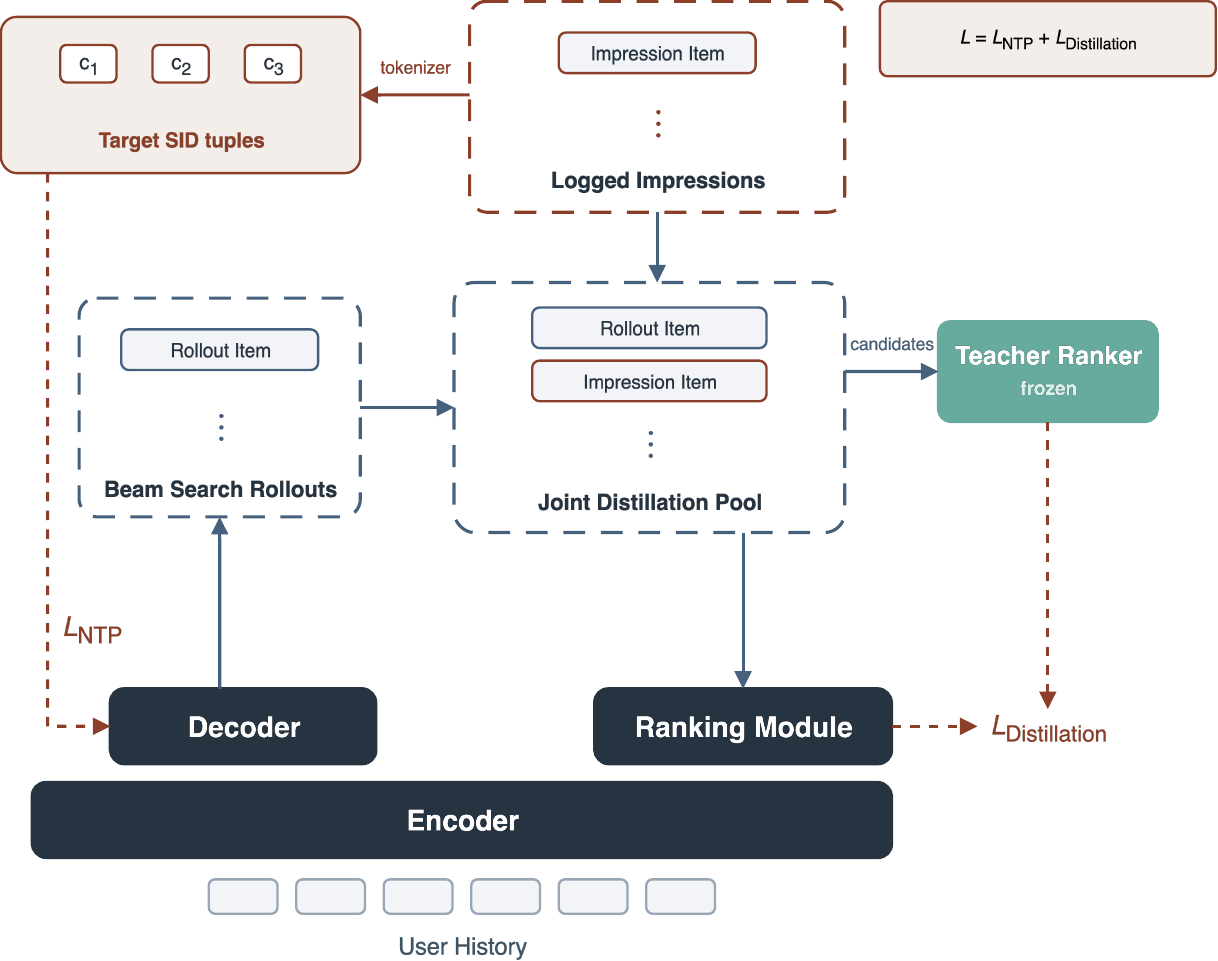}
  \caption{Joint training of the decoder ($\mathcal{L}_\text{NTP}$)
  and the Ranking Module ($\mathcal{L}_\text{Distillation}$, regression
  toward the frozen Teacher Ranker) on the shared encoder.}
  \label{fig:sona-distill}
\end{figure}

\Sona{} is trained with a joint generation-and-ranking objective
(\Cref{fig:sona-distill}). The decoder is supervised by next-token
prediction (NTP) on the \emph{positive} impressions of each request only ---
the items the user engaged with, mapped to their Semantic ID tuples
(\Cref{sec:train:data}). The encoder runs once per request, and every
positive is decoded as an independent teacher-forced sequence that
cross-attends to the shared representation. The Ranking Module is
supervised by distillation from the frozen Teacher Ranker
(\Cref{sec:teacher:arch}): the teacher exposes \(n\) engagement
scores per candidate, and the Ranking Module regresses a matching head
onto each with an element-wise mean absolute error. Both objectives
back-propagate into the shared encoder.

\paragraph{Distillation sources.} The distillation objective draws
(user, candidate) pairs from two complementary sources: current-decoder
rollouts and logged impressions. For each user history, beam search
generates \(B\) SID tuples using the current decoder. The generated SIDs
are resolved into catalog items; successfully resolved items form the
rollout candidate set \(\mathcal{B}\), while unresolved entries are
excluded from Ranking Module scoring and loss. The rollout set is
complemented with the request's logged impressions \(\mathcal{I}\),
which broaden teacher supervision to items exposed in production
traffic. Writing \(r^{h}_{c}\) for the teacher's head-\(h\) score
of candidate \(c\) and \(\hat r^{\,h}_{c}\) for the Ranking Module's,
each source contributes a per-head mean absolute error,
\[
  \mathcal{L}_\text{rollout}
    = \sum_{h=1}^{n}
      \frac{1}{\lvert\mathcal{B}\rvert}
      \sum_{c\in\mathcal{B}}
      \bigl\lvert\,\hat r^{\,h}_{c} - r^{h}_{c}\,\bigr\rvert ,
  \qquad
  \mathcal{L}_\text{impression}
    = \sum_{h=1}^{n}
      \frac{1}{\lvert\mathcal{I}\rvert}
      \sum_{c\in\mathcal{I}}
      \bigl\lvert\,\hat r^{\,h}_{c} - r^{h}_{c}\,\bigr\rvert .
\]
The joint loss sums the generative and the two distillation terms,
\[
  \mathcal{L} = \mathcal{L}_\text{NTP}
              + \mathcal{L}_\text{rollout}
              + \mathcal{L}_\text{impression}.
\]

\paragraph{Combined ranking score.} At serving, the \(n\) distilled
head scores are collapsed into one scalar by the fixed weighted
combiner of \Cref{sec:arch:musicgryphon}. This score reorders the beam
candidates; it enters no training loss.

\paragraph{Training stages.} \emph{Initial training} starts from random
initialization and consumes the dataset chronologically in day-level
chunks --- days in order, requests shuffled within a day --- under a
linear warmup-and-decay learning-rate schedule. \emph{Online training}
then consumes requests as they arrive (\Cref{sec:infra:online}), at a
constant learning rate, the terminal rate of the decay.

\paragraph{Optimization.} Training uses AdamW with weight decay and
gradient clipping; optimizer and schedule hyperparameters are listed in
\Cref{app:hyperparameters}. FSDP shards the trainable parameters, while
the frozen teacher stays replicated in bfloat16 on every rank, adding
no all-gather cost.

\section{Teacher Ranker}
\label{sec:teacher}

\subsection{Architecture}
\label{sec:teacher:arch}

\begin{figure}[t]
  \centering
  \includegraphics[width=0.7\linewidth]{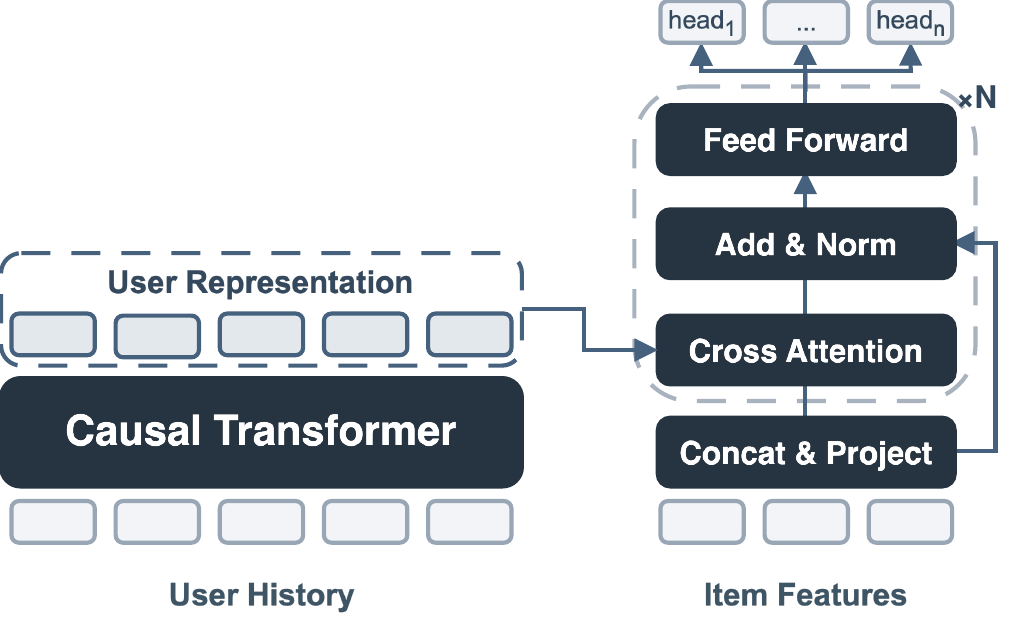}
  \caption{Teacher Ranker architecture: a history encoder over the
  typed event sequence (left) and a candidate scorer that cross-attends
  to its hidden states (right).}
  \label{fig:sona-reward}
\end{figure}

The teacher ranker is a large ranker without hand-engineered features, used
as the distillation teacher of the Ranking Module
(\Cref{sec:train:rescorer}). Freed from the latency and throughput budgets
of the final model, it is larger and attends the full user history. It
surpasses the production ranker in online experiments
(\Cref{sec:eval:online}).
\Cref{fig:sona-reward} shows the architecture.

The teacher combines a deep user-history encoder over the typed event sequence
with a candidate scorer. As in \Sona{}'s Ranking Module
(\Cref{sec:arch:encoder,sec:arch:musicgryphon}), the scorer largely follows the
decoder architecture: it takes a candidate embedding as input and refines it
against the encoder's hidden states through multi-layer cross-attention
\citep{hstu2024, grank2026}. Two differences separate it from the served pair.
First, its encoder is \emph{causal}: every position carries the user
state at that point of the timeline, which unlocks autoregressive
training and timeline-packed supervision
(\Cref{sec:teacher:training}). Second, it
does not use History Compression --- the deep stack attends the full
user history at every layer.

\paragraph{Heads.} A set of engagement heads reads from the scorer's
output, each an MLP that narrows to one scalar. One head supplies the
primary ranking signal, trained to order events by a graded engagement
target; the others are auxiliary feedback heads, restored to calibrated
logits by a learned per-head scale and bias
(\Cref{sec:teacher:training}). The complete model card is given in \Cref{tab:hparams:reward}.

\subsection{Dataset Design}
\label{sec:teacher:data}

The Teacher Ranker follows a two-stage initial recipe --- pre-training with
next-item prediction (NIP), then multi-head ranking fine-tuning --- followed by
continuous refresh (\Cref{sec:teacher:training}). Both initial stages and the
refresh runs draw on the same dataset shape. The dataset is timeline-level: its unit is a span of one
user's chronological event timeline, carrying the same logged
attributes as \Cref{tab:arch:backbone:features}. Pre-training samples
consist of the timeline alone, and every event in it is a next-item
target. Fine-tuning samples additionally carry the impressions served
along the timeline, with their graded engagement feedback.
Refresh samples keep the fine-tuning format and are built
continuously from newly logged sessions, restricted to the surface the
model serves.

\paragraph{Autoregressive consumption.} Each sample is consumed with
all of its targets at once. The causal encoder provides, at every
position, the user state at that point of the timeline, and the
cross-attention scorer reads each target through a custom mask that
restricts it to the history preceding its anchor
(\Cref{sec:teacher:training}). One encoder pass therefore supervises
every target the timeline carries, where the request-level dataset of
\Cref{sec:train:data} runs the encoder once per request.

\paragraph{Data span.} Packing a timeline's worth of targets into one
encoder pass sets the affordable window: the teacher is trained on a
\emph{year} of target events, where the joint run of
\Cref{sec:train:data} consumes several weeks. Compressing that year
into the teacher's weights, and distilling it densely into the Ranking
Module, is the design idea of \Cref{sec:philosophy}.

\subsection{Training Framework}
\label{sec:teacher:training}

The Teacher Ranker is trained autoregressively over the user's
chronological engagement-event sequence, using the recipe of
\Cref{sec:teacher:data}: NIP pre-training learns the user encoder,
multi-head ranking fine-tuning specializes it to the signal the Ranking
Module consumes, and continuous refresh keeps the fine-tuned teacher current.
The contribution of the pre-training stage
is ablated in \Cref{tab:eval:reward}.

\paragraph{Pre-training.} The history encoder is pre-trained on
next-item prediction, following the autoregressive recipe of Argus
\citep{argus2025}: every position of the timeline predicts the next
engagement from its causal user state. The objective is an InfoNCE
contrastive loss over in-batch negatives mixed with a uniformly
sampled tail-negative pool, with LogQ correction \citep{logq2019} for
the popularity bias of in-batch sampling. The stage keeps the encoder
parameters; the prediction head is discarded.

\paragraph{Fine-tuning.} The pre-trained encoder is warm-started, the
candidate scorer is attached with random initialization, and the model
is trained as a multi-objective ranker. The primary objective is
\emph{pairwise}. Each impression is mapped to a graded engagement
scale --- \(\text{like} > \text{play} > \text{skip} >
\text{dislike}\) --- and the pairwise head is trained to order
temporally adjacent impressions with different grades by a logistic
loss on its score difference. Over the set \(\mathcal{P}\) of ordered
adjacent pairs \((i, j)\) in which \(i\) carries the higher grade,
\begin{equation}
  \mathcal{L}_\text{pair}
    = - \frac{1}{\lvert \mathcal{P} \rvert}
      \sum_{(i, j) \in \mathcal{P}}
      \log \sigma\!\left(
        \mathrm{score}_{\mathrm{pair}}(i)
        - \mathrm{score}_{\mathrm{pair}}(j)
      \right).
  \label{eq:train:reward:pair}
\end{equation}
Pairs form both within one recommendation request and across the
boundary of two adjacent ones, and only impressions served in
recommendation requests are eligible: fine-tuning stays on the
distribution the ranker is used on. The auxiliary objectives are
\emph{pointwise}: each remaining head fits one binary feedback target
with binary cross-entropy \(\mathcal{L}_{\mathrm{BCE}}^{k}\)
on its calibrated logit. The heads are trained jointly against the sum
of the pairwise and pointwise losses,
\[
  \mathcal{L}_\text{teacher}
    = \mathcal{L}_\text{pair}
    + \sum_{k} \mathcal{L}_{\mathrm{BCE}}^{k}.
\] Each impression group attends the history preceding
its anchor through the custom cross-attention mask of
\Cref{sec:teacher:data}, implemented as a dedicated Triton kernel
\citep{triton2019}. Fine-tuning extends the attended history well
beyond the pre-training length, so the scorer sees the long-range
context available at scoring time.

\paragraph{Continuous refresh.} After fine-tuning, the teacher is rolled
forward to the present and then refreshed daily on newly logged
sessions. These continuation runs narrow the target pairs to the
recommendation surface whose generative model the teacher supervises,
specializing its signal to that surface.

\paragraph{Optimization.} All stages use Adam in two parameter
groups --- encoder and heads --- with a linear-warmup schedule,
following \citet{argus2025}, under bfloat16 autocast. Sequence
lengths, negative counts, and learning rates are listed in the model
card (\Cref{tab:hparams:reward}).

\section{Training and Inference Infrastructure}
\label{sec:infra}

\subsection{Training Optimizations}
\label{sec:infra:training}

\Sona{} trains generation and ranking jointly and evaluates the frozen teacher on
the generated candidates in the same step. We co-locate all three computations
on one set of GPU workers.

\paragraph{Keep the data path off the critical path.} After network fetch and
CPU preprocessing, the input pipeline buffers prepared batches for the GPU to
consume. Catalog item hashes are computed by
the asynchronous input pipeline while training proceeds. Data preparation,
hash computation, and transfer can therefore overlap GPU model execution.

\paragraph{Optimize GPU execution.} We use bfloat16 mixed precision to shorten
the processing time of each batch. FSDP \citep{fsdp2023} shards the trainable
model state. FlashAttention \citep{dao2022} supplies efficient attention
kernels, and \texttt{torch.compile} optimizes the dense model graph.

\paragraph{Preserve numerical stability.} QK normalization
\citep{qknorm2020} bounds attention logits, and gradient clipping limits
outlier updates. We also tune the learning-rate schedule carefully, including
warmup before the peak rate and decay during training. These controls are
important when combining mixed precision, long histories, and the joint
generation--ranking objective.

\subsection{Online Training Infrastructure}
\label{sec:infra:online}

\begin{figure}[t!]
  \centering
  \includegraphics[width=\linewidth]{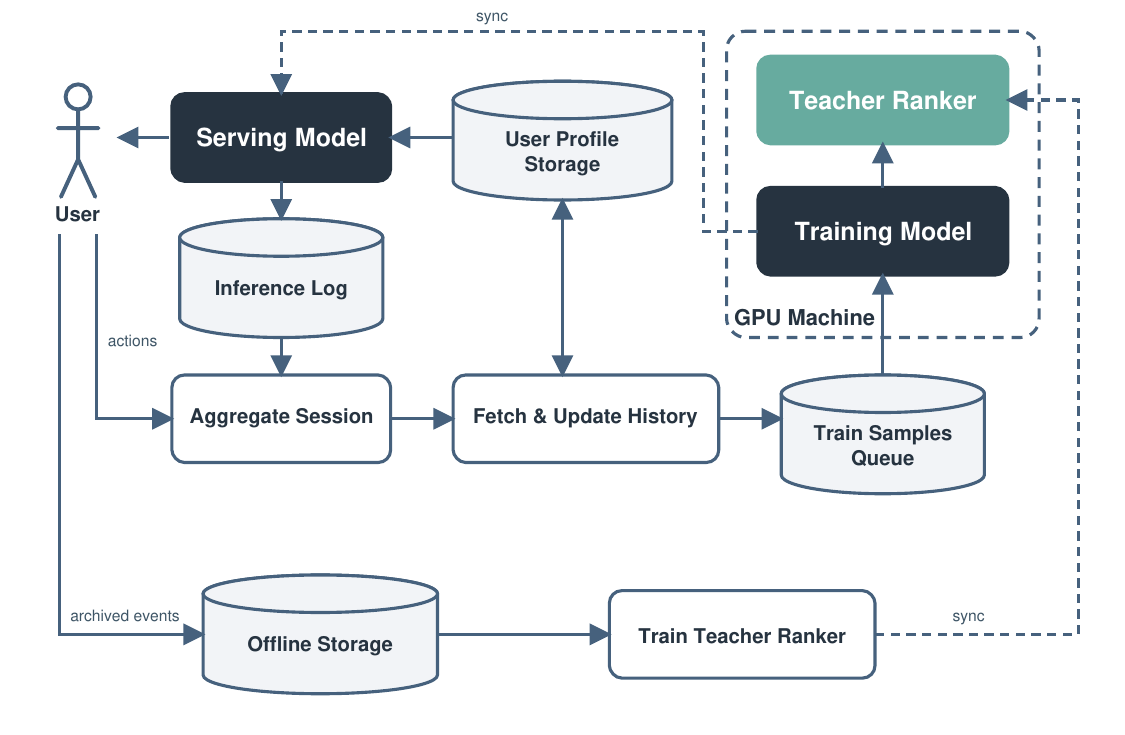}
  \caption{Online-training infrastructure for the final \Sona{} configuration. In the
  diagram, ``Teacher Ranker'' is the frozen distillation teacher,
  ``Train Teacher Ranker'' its training pipeline, ``Training
  Model'' / ``Serving Model'' the trainer and serving copies of
  \Sona{}, and ``Inference Log'' the per-request serving record that
  the session aggregator joins by session identifier.}
  \label{fig:sona-online}
\end{figure}

\Cref{fig:sona-online} shows the online-training data path. Updates to the
served model come from a continuous loop: user actions stream into a
real-time event-processing system, are aggregated per user into
sessions, joined with the inference-time context, and emitted as
training samples; a GPU trainer consumes those samples from a message
queue and updates the model through the joint objective of
\Cref{sec:train:pretrain} --- next-token prediction and Ranking Module
distillation; the updated model is checkpointed and
synced back to serving every \NumModelSyncPeriod{}. The Teacher Ranker
runs its own training pipeline
(\Cref{sec:teacher:training}) and is refreshed into the trainer every
\NumRewardRefreshCadence{}. Measured from when a user event is emitted until
the model update trained on it reaches serving, the end-to-end latency is
\NumOnlineLagMedian{} at the median and
\NumOnlineLagPNinetyNine{} at the 99th percentile. The
session-attribution window of \NumAttributionWindow{} described below
and the deployment cadence of \NumModelSyncPeriod{} account for most
of the median budget; event processing, queueing, training, and artifact
upload occupy the remainder.

\paragraph{Session aggregation.} The event stream is partitioned by
user and consumed by a stream processor whose per-user state is
backed by a sharded distributed key--value store. The processor groups
consecutive events into \emph{sessions} closed by an idle gap of a
few minutes and by a minimum attribution window of
\NumAttributionWindow{}; the attribution window also serves as a correctness
buffer that lets the inference log catch up to the events that
triggered it (below). Because the stream processor wakes a worker
only when a new message arrives, session timeouts are driven by
internal timer messages that the worker enqueues for itself.

\paragraph{Traffic filtering.} Emitted sessions pass through several
gates: standard fraud and bot-activity filters; a history-length gate
that keeps the gradient from being dominated by very short histories;
and a model blacklist that drops traffic served by probe or
evaluation models, which would otherwise close a feedback loop with
the training data.

\paragraph{Concurrent sample streams.} Session aggregation is
configured per model. Several models are registered at a time, each
with its own session definition, traffic gates, and sample format,
and each emitting its own stream.
Several online updates therefore run side by side over one pass of
the user events --- one feeding the deployed model, the others
feeding candidate variants under evaluation.

\paragraph{User profile storage.} The user history that serving and
sample construction both read is held in two stores behind one
logical profile. A key--value store holds the long history and is
rebuilt on a batch cadence, so its tail trails live traffic. A
real-time profile, written directly from the event stream, holds the
last \NumProfileRealtimeWindow{} of events, capped at
\NumProfileRealtimeCap{} events per user. A read merges the two by
timestamp and deduplicates the overlap, resolving collisions in
favor of the batch side. The window on the real-time side is what
keeps the split worthwhile: the stream processor holds and updates a
bounded tail per user instead of the full \NumCompressMaxLen{}-event
history, which is what keeps its state size and write load
affordable at the request rate of the surface.

\paragraph{History join via the inference service.} A training sample
is meaningful only if the user history it shows the model matches the
history the inference saw when the candidate was produced; otherwise
the gradient is computed against a state the model never observed.
Two design choices make this tractable. First, the inference service
writes an \emph{inference log} entry to a replicated key--value store
at the moment of serving, keyed by the request identifier and
carrying the history-cutoff timestamp at inference time; the session
aggregator joins this entry by session identifier and stamps the
sample with the cutoff. Second, the user history itself is not
propagated through the event stream. The aggregated session is posted
back to the inference service, which re-reads the profile from the
two stores and rebuilds the sample through the same computation it
runs at serving time (\Cref{sec:infra:inference}).
Carrying history inline in every event would inflate the streaming
bandwidth by orders of magnitude relative to the rate of training
samples, whereas a rebuild from storage stays at the rate of emitted
samples.

\paragraph{Update loop.} The emitted training samples land in a
message queue with at-least-once delivery and single-digit-second
read latency. The GPU trainer consumes from the queue and applies
the joint objective of \Cref{sec:train:pretrain} --- the
next-token-prediction and Ranking Module distillation terms --- using the
current model as the rollout source for distillation.
Updated models are uploaded to a model-artifact store every
\NumModelSyncPeriod{} and pulled by the serving fleet on the same
cadence.

\subsection{Inference Optimizations}
\label{sec:infra:inference}

The served \Sona{} model contains the shared encoder, autoregressive decoder,
and Ranking Module. In the final production configuration, the frozen teacher
remains outside the serving path and is evaluated only by the trainer;
temporary teacher-ranked A/B treatments used a separate experimental path
(\Cref{sec:eval:online}). Generation and item-level ranking are executed by the
same GPU worker and share the encoded user state.
With the optimizations described below, the served model reaches a model FLOPs
utilization (MFU) of \NumInferenceMFU{} during inference.

\paragraph{Sample construction.} A separate CPU inference service assembles the
model input: it reads the user profile (\Cref{sec:infra:online}) and
computes the event attributes of \Cref{tab:arch:backbone:features}
through a feature store, then sends the finished sample to the GPU
fleet. The same feature-store processing is used at every stage ---
initial training, online training, and inference --- so event
attributes are computed identically on all three paths.

\paragraph{NVIDIA Triton Inference Server.} NVIDIA Triton Inference Server
separates request preparation from model execution. An asynchronous CPU batcher groups variable-length user histories,
truncates and pads them to fixed input shapes, and dispatches each packed batch
to a GPU worker. Fixed shapes allow the worker to capture the complete model
path as a CUDA graph during initialization. Steady-state
inference copies new inputs into the captured graph and replays it, reducing
framework and kernel-launch overhead across the complete request.

\paragraph{Model execution.} The history encoder runs once per packed batch.
Its hidden states condition every Semantic ID decoding step and are reused by
the Ranking Module. Intermediate candidates remain on GPU throughout
generation and ranking. The model subgraphs are compiled with
\texttt{torch.compile}, and the main computation uses bfloat16.

\paragraph{Beam-search selection.} With a beam width of
\NumInferenceBeamSize{} and \NumCodebookSize{} logits at each Semantic ID
level, top-$k$ selection is a material decoding cost. We apply the radix-based
selector of \citet{flashinfer2025} within each parent beam, then merge the
retained candidates into the next global beam. Catalog-prefix constraints
remove invalid extensions, and the key--value (KV) cache follows the selected
parent hypotheses.

\paragraph{Decoder attention.} Large-beam decoding produces many independent
one-token queries with key--value sequences of at most \NumSemIDLevels{}
positions, a geometry poorly matched to the default FlashAttention tile sizes
\citep{dao2022}. We therefore use a decoder self-attention kernel with tile
sizes tuned for this geometry, substantially reducing computation overhead.
The decoder maintains a KV cache across generation steps.

\section{Evaluation}
\label{sec:evaluation}

We evaluate four questions. First, how do tokenizer choice, model capacity, and
data volume affect candidate generation? Second, how do pre-training, history
length, and scorer depth affect the Teacher Ranker's ranking objective? Third,
which training and architecture choices let the unified model preserve both
target-track retrieval and the teacher's ordering? Finally, does the selected
single-model configuration improve engagement online?
\Cref{tab:eval:configurations} maps each question to the model under test and
the reported evidence.

\begin{table}[H]
  \centering
  \footnotesize
  \begin{tabularx}{\linewidth}{@{}
    >{\raggedright\arraybackslash}p{0.22\linewidth}
    >{\raggedright\arraybackslash}X
    >{\raggedright\arraybackslash}X@{}}
    \toprule
    \textbf{Evaluation question} & \textbf{Model under test}
      & \textbf{Reported evidence} \\
    \midrule
    Candidate generation
      & NTP-only encoder--decoder
      & Target-track Recall@\(k\); train NTP loss at matched exposure for the
        capacity sweep \\
    \addlinespace
    Teacher Ranker design
      & Teacher Ranker variants under a shared offline protocol
      & Weighted pair accuracy (WPA) across pre-training, history-length, and
        scorer-depth ablations \\
    \addlinespace
    Unified-model transfer
      & Encoder--decoder and Ranking Module trained jointly with NTP and
        distillation; teacher used only for distillation targets and reference
        scores
      & Target-track Recall@\(k\), Teacher Recall@\(k\), and WPA \\
    \addlinespace
    Online performance
      & Intermediate treatments and the complete \Sona{} serving configuration
      & Relative changes in engagement metrics against the control arm from
        the same experiment \\
    \bottomrule
  \end{tabularx}
  \caption{Scope of the evaluation.}
  \label{tab:eval:configurations}
\end{table}

\subsection{Evaluation Protocol and Metrics}
\label{sec:eval:protocol}

\paragraph{Target-track Recall.} Let \(n_i\) be the number of target tracks for
request \(i\), and let \(m_i(k)\) be the number of those targets found among the
first \(k\) resolved track predictions. Target-track Recall@\(k\) is the
macro-average of \(\min\{m_i(k), k\}/\min\{n_i, k\}\) over held-out requests
with at least one target.
The model-capacity sweep instead reports train NTP loss at matched target
exposure; lower train loss is better.

\paragraph{Teacher Recall.} Teacher Recall@\(k\) evaluates
ranking fidelity conditional on candidate generation: for a held-out request
\(r\), the decoder produces a candidate pool \(C_r\), and the Ranking Module
and teacher rank the same pool. If \(\mathcal{R}\) is the set of evaluated
requests and their scalar scores are \(s_{\mathrm{RM}}\) and
\(s_{\mathrm{teacher}}\), respectively, then
\[
\mathrm{TeacherRecall@}k
=
\frac{1}{|\mathcal{R}|}
\sum_{r\in\mathcal{R}}
\frac{1}{k}
\left|
  \mathrm{Top}_k(s_{\mathrm{RM}}^{(r)};C_r)
  \cap
  \mathrm{Top}_k(s_{\mathrm{teacher}}^{(r)};C_r)
\right|,
\]
where the intersection is computed within the common candidate pool. When a
configuration has no Ranking Module, decoder likelihood supplies its ordering.

\paragraph{Weighted pair accuracy.} Weighted pair accuracy measures
engagement-based ordering on held-out pairs of temporally adjacent impressions
with different target weights. If \(t_i\) is the predefined target weight of
impression \(i\) and \(\mathcal{P}\) contains pairs with \(t_i>t_j\), then
\begin{equation}
  \operatorname{WeightedPairAccuracy}
  =
  \frac{
    \sum_{(i,j)\in\mathcal{P}}
      (t_i-t_j)\,\mathbf{1}\!\left\{
        \mathrm{score}_{\mathrm{pair}}(i)
        > \mathrm{score}_{\mathrm{pair}}(j)
      \right\}
  }{
    \sum_{(i,j)\in\mathcal{P}} (t_i-t_j)
  }.
  \label{eq:eval:weighted-pair-accuracy}
\end{equation}
Pairs farther apart in target weight therefore contribute more. All three
offline metrics are higher-is-better.

\subsection{Candidate Generation}
\label{sec:eval:retrieval}

These experiments evaluate an NTP-only encoder--decoder, without the Ranking
Module or teacher. The tokenizer and data-volume ablations use Target-track
Recall@\(k\) on held-out requests; the capacity sweep compares train NTP loss
at matched target exposure. We vary the tokenizer, model size, and training-data
volume.

\paragraph{Semantic tokenizer.}
\label{sec:eval:tokenizer}

Each tokenizer maps a dense track representation to a three-code Semantic ID
tuple using residual K-means. We vary codebook size and item representation
independently, regenerating the IDs and retraining the retrieval model for each
setting. CLMR, a contrastively trained music-audio encoder \citep{clmr2021},
provides the audio-only baseline. Holding CLMR embeddings fixed, we compare
three-level quantizers with \(8{,}192\) or \(32{,}000\) entries per codebook.
Holding the quantizer configuration at three \(32{,}000\)-entry codebooks, we
compare raw CLMR, CLMR with an InfoNCE-trained residual projection on
collaborative pairs and the Qwen2.5-Omni audio-and-metadata representation with
transformer-based collaborative refinement from \Cref{sec:arch:tokenizer}.

\begin{table}[H]
  \centering
  \footnotesize
  \begin{subtable}[t]{0.29\linewidth}
    \centering
    \begin{tabular}{lc}
      \toprule
      \textbf{Codebook size} & \textbf{Recall@1000} \\
      \midrule
      $3 \times 8{,}192$  & $0.8036$ \\
      $3 \times 32{,}000$ & $\mathbf{0.8111}$ \\
      \bottomrule
    \end{tabular}
    \caption{Codebook size.}
    \label{tab:eval:ablation:codebook}
  \end{subtable}
  \hfill
  \begin{subtable}[t]{0.69\linewidth}
    \centering
    \setlength{\tabcolsep}{3pt}
    \begin{tabular}{lccc}
      \toprule
      \textbf{Item representation} &
      \textbf{Recall@10} &
      \textbf{Recall@100} &
      \textbf{Recall@1000} \\
      \midrule
      CLMR audio embedding
        & $0.1848$ & $0.4712$ & $0.8111$ \\
      \shortstack[l]{CLMR + collaborative\\projection}
        & $0.2005$ & $0.4873$ & $0.8139$ \\
      \shortstack[l]{Qwen2.5-Omni + collaborative\\refinement}
        & $\mathbf{0.2171}$ & $\mathbf{0.5362}$ & $\mathbf{0.8524}$ \\
      \bottomrule
    \end{tabular}
    \caption{Item representation.}
    \label{tab:eval:tokenizer-approaches}
  \end{subtable}
  \caption{Semantic-tokenizer ablations. Each subtable varies one component
  and holds the other fixed. All variants use the same held-out set.}
  \label{tab:eval:tokenizer-ablations}
\end{table}

We select the \(3 \times 32\mathrm{k}\) codebook and refined Qwen2.5-Omni
representation, which provides the strongest retrieval result in this
ablation.

\paragraph{Model size.}
\label{sec:eval:model-size}

The model-size sweep varies encoder--decoder capacity while fixing the layer
allocation, data window, objective, and \NumCompressMaxLen{}-event histories.
We report train NTP loss against cumulative packed-target exposure. The Small,
Medium, and \(2\times\) Medium configurations contain
\NumScalingSmallCoreParams{},
\NumScalingMediumCoreParams{}, and \NumScalingLargeCoreParams{} transformer-core
parameters, respectively. Their total learnable counts, including embeddings
and output heads, are \NumScalingSmallTotalLearnableParams{},
\NumScalingMediumTotalLearnableParams{}, and
\NumScalingLargeTotalLearnableParams{}.

\begin{figure}[H]
  \centering
  \begin{tikzpicture}
    \begin{axis}[
      width=0.94\linewidth,
      height=0.48\linewidth,
      xlabel={Cumulative packed targets (billions)},
      ylabel={Train NTP loss},
      xmin=0.30,
      xmax=2.35,
      ymin=1.88,
      ymax=2.34,
      xtick distance=0.25,
      ytick distance=0.10,
      grid=major,
      major grid style={draw=gray!20},
      tick label style={font=\small},
      label style={font=\small},
      legend style={
        at={(0.5,1.03)},
        anchor=south,
        draw=none,
        legend columns=3,
        font=\small
      },
    ]
      \addplot[
        black!65,
        densely dotted,
        line width=1.2pt
      ] table[
        x=targets_billions,
        y=small,
        col sep=comma
      ] {figures/sona_model_scaling_train_loss.csv};
      \addlegendentry{Small (\NumScalingSmallCoreParams{} core)}

      \addplot[
        blue!70!black,
        dashed,
        line width=1.2pt
      ] table[
        x=targets_billions,
        y=medium,
        col sep=comma
      ] {figures/sona_model_scaling_train_loss.csv};
      \addlegendentry{Medium (\NumScalingMediumCoreParams{} core)}

      \addplot[
        red!75!black,
        solid,
        line width=1.2pt
      ] table[
        x=targets_billions,
        y=large,
        col sep=comma
      ] {figures/sona_model_scaling_train_loss.csv};
      \addlegendentry{\(2\times\) Medium (\NumScalingLargeCoreParams{} core)}
    \end{axis}
  \end{tikzpicture}
  \caption{Train NTP loss of the Small, Medium, and \(2\times\) Medium
  configurations at matched cumulative packed-target exposure.}
  \label{fig:eval:model-size}
\end{figure}
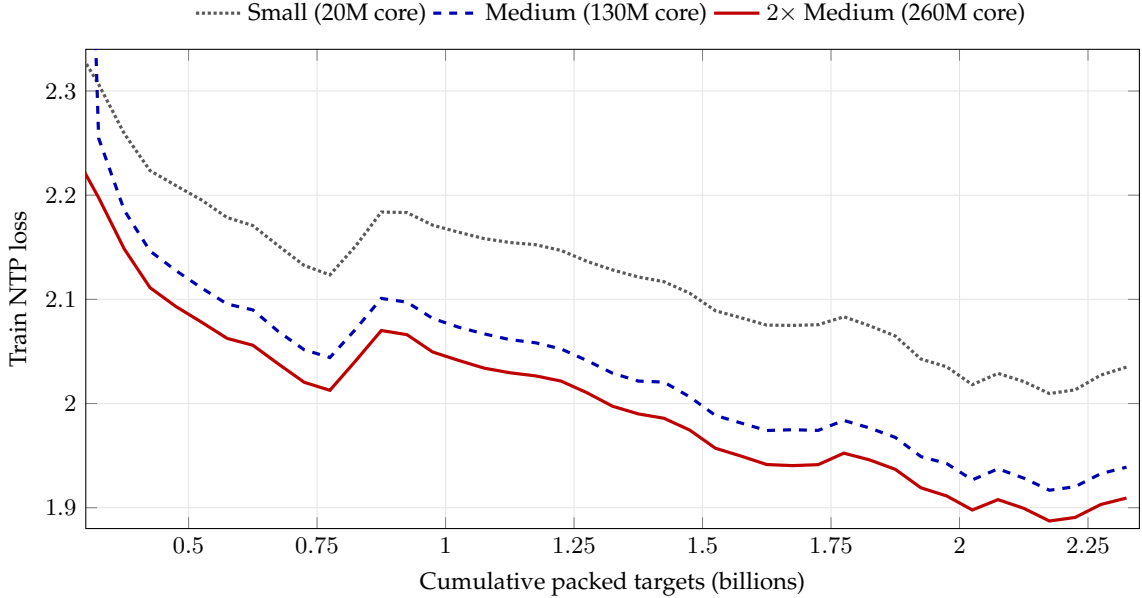

Increasing capacity consistently lowers train NTP loss over the measured
exposure range.

\paragraph{Training-data volume.} We next fix the Medium backbone,
\NumBaselineHistoryLength{}-event histories, tokenizer, held-out set, and
optimization recipe. Each run makes one chronological pass over a longer log
window.

\begin{table}[H]
  \centering
  \small
  \begin{tabular}{lcccc}
    \toprule
    \textbf{Training window} &
    \textbf{Train loss} &
    \textbf{Recall@10} &
    \textbf{Recall@100} &
    \textbf{Recall@1000} \\
    \midrule
    \NumDataScalingOneWeekWindow{} &
    \NumDataScalingOneWeekTrainLoss{} &
    \NumDataScalingOneWeekRecallTen{} &
    \NumDataScalingOneWeekRecallHundred{} &
    \NumDataScalingOneWeekRecallThousand{} \\
    \NumDataScalingTwoWeekWindow{} &
    \NumDataScalingTwoWeekTrainLoss{} &
    \NumDataScalingTwoWeekRecallTen{} &
    \NumDataScalingTwoWeekRecallHundred{} &
    \NumDataScalingTwoWeekRecallThousand{} \\
    \NumDataScalingFourWeekWindow{} &
    \NumDataScalingFourWeekTrainLoss{} &
    \NumDataScalingFourWeekRecallTen{} &
    \NumDataScalingFourWeekRecallHundred{} &
    \NumDataScalingFourWeekRecallThousand{} \\
    \NumDataScalingEightWeekWindow{} &
    \NumDataScalingEightWeekTrainLoss{} &
    \NumDataScalingEightWeekRecallTen{} &
    \NumDataScalingEightWeekRecallHundred{} &
    \NumDataScalingEightWeekRecallThousand{} \\
    \bottomrule
  \end{tabular}
  \caption{Final encoder--decoder backbone train loss and recall as the unique
  chronological training-data window grows. Each run makes one pass
  over its window.}
  \label{tab:eval:data-volume-scaling}
\end{table}

The point estimates improve monotonically as the one-pass training window
grows, including from \NumDataScalingFourWeekWindow{} to
\NumDataScalingEightWeekWindow{}. Together with the separated capacity curves,
the sweep shows no observed plateau within the measured ranges.

\subsection{Teacher Ranker Design Ablations}
\label{sec:eval:teacher}

These experiments compare Teacher Ranker variants using the held-out
engagement-pair protocol and WPA defined above. They isolate choices within the
teacher --- initial pre-training, attended history length, and candidate-scorer
depth.

\paragraph{Pre-training.} We compare the two-stage initial recipe with direct
fine-tuning from random initialization. Both variants use an \(8\text{k}\)-event
history and a six-layer candidate scorer; the only change is whether
next-item-prediction pre-training precedes ranking fine-tuning.

\begin{table}[H]
  \centering
  \small
  \begin{tabular}{lc}
    \toprule
    \textbf{Configuration}             & \textbf{Weighted pair accuracy} \\
    \midrule
    Teacher, no pre-training           & \NumRewardWeightedPairNoPretrain{} \\
    Teacher, two-stage initial recipe (selected)
                                       & \textbf{\NumRewardWeightedPairPretrained{}} \\
    \bottomrule
  \end{tabular}
  \caption{Teacher Ranker pre-training ablation at fixed \(8\text{k}\)-event
  history and six-layer candidate scorer. The first row removes only the
  next-item-prediction pre-training stage of \Cref{sec:teacher:training}.}
  \label{tab:eval:reward}
\end{table}

The gain establishes NTP pre-training as part of the selected Teacher Ranker
recipe.

\paragraph{History length and scorer depth.} We next vary both axes while
holding the remaining teacher configuration fixed.

\begin{table}[H]
  \centering
  \small
  \begin{tabular}{llc}
    \toprule
    \textbf{History} & \textbf{Scorer depth}
                     & \textbf{Weighted pair accuracy} \\
    \midrule
    \multirow{5}{*}{\(2\text{k}\)}
      & 1 & \NumRewardWeightedPairHistoryTwoKDepthOne{} \\
      & 2 & \NumRewardWeightedPairHistoryTwoKDepthTwo{} \\
      & 4 & \NumRewardWeightedPairHistoryTwoKDepthFour{} \\
      & 6 & \NumRewardWeightedPairHistoryTwoKDepthSix{} \\
      & 8 & \NumRewardWeightedPairHistoryTwoKDepthEight{} \\
    \midrule
    \multirow{5}{*}{\(8\text{k}\)}
      & 1 & \NumRewardWeightedPairHistoryEightKDepthOne{} \\
      & 2 & \NumRewardWeightedPairHistoryEightKDepthTwo{} \\
      & 4 & \NumRewardWeightedPairHistoryEightKDepthFour{} \\
      & 6 (selected) & \NumRewardWeightedPairPretrained{} \\
      & 8 & \textbf{\NumRewardWeightedPairHistoryEightKDepthEight{}} \\
    \bottomrule
  \end{tabular}
  \caption{Teacher Ranker WPA point estimates as candidate-scorer depth and
  attended history length vary on the same held-out engagement-label set.}
  \label{tab:eval:reward:depth}
\end{table}

Longer attended histories provide consistent gains across scorer depths. We
use the six-layer, \(8\text{k}\)-event configuration in subsequent distillation
and limited-traffic experiments.

\subsection{Unified Generation and Ranking}
\label{sec:eval:unified}

With the teacher configuration fixed, these offline experiments evaluate the
encoder, decoder, and Ranking Module trained jointly with NTP and distillation.
The teacher supplies distillation targets and evaluation reference scores but
is not part of the evaluated model. We report the metrics defined in
\Cref{sec:eval:protocol} while varying the distillation loss, candidate source,
training rollout beam size, Ranking Module depth, and history encoding.

\paragraph{Experimental sequence.} All variants use the components of
\Cref{sec:architecture} and a \NumBaselineSFTWeeks{}-week training window. We
report encoder, decoder, and Ranking Module layer counts separately in the
\(E\), \(D\), and \(R\) columns of
\Cref{tab:eval:scorer-history}. A dash in the \(R\) column denotes
decoder-likelihood ordering without a Ranking Module.
\Cref{tab:eval:unified-ablations}
varies one factor at a time around a shared reference: MAE distillation,
rollouts plus logged impressions, a training rollout beam of $32$,
\NumBaselineHistoryLength{}-event histories, and a
model with \(E=\NumBaselineEncoderLayers{}\),
\(D=\NumBaselineDecoderLayers{}\), and
\(R=\NumItemScorerOneLayerDepth{}\). The daggered entry in each block is
therefore the same reference run. Every variant is evaluated with the fixed
serving beam \(B_{\mathrm{eval}}=\NumInferenceBeamSize{}\). We then increase
\(R\) to \NumBaselineRankingModuleLayers{} and use that architecture for the
history ablations.

\begin{table}[H]
  \centering
  \footnotesize
  \setlength{\tabcolsep}{4pt}
  \begin{tabular}{@{}lcccc@{}}
    \toprule
    \textbf{Configuration}
      & \textbf{\shortstack{Target-track\\Recall@1000}}
      & \textbf{\shortstack{Teacher\\Recall@10}}
      & \textbf{\shortstack{Teacher\\Recall@100}}
      & \textbf{WPA} \\
    \midrule

    \multicolumn{5}{l}{\textit{Distillation loss}} \\
    MAE$^{\dagger}$
      & $\NumDistillMAERecallThousand{}$
      & $\NumDistillMAETeacherRecallTen{}$
      & $\mathbf{\NumDistillMAETeacherRecallHundred{}}$
      & $\NumDistillMAEWeightedPair{}$ \\
    MSE
      & $\NumDistillMSERecallThousand{}$
      & $\mathbf{\NumDistillMSETeacherRecallTen{}}$
      & $\NumDistillMSETeacherRecallHundred{}$
      & $\mathbf{\NumDistillMSEWeightedPair{}}$ \\
    Huber
      & $\mathbf{\NumDistillHuberRecallThousand{}}$
      & $\NumDistillHuberTeacherRecallTen{}$
      & $\NumDistillHuberTeacherRecallHundred{}$
      & $\NumDistillHuberWeightedPair{}$ \\
    KL
      & $\NumDistillKLRecallThousand{}$
      & $\NumDistillKLTeacherRecallTen{}$
      & $\NumDistillKLTeacherRecallHundred{}$
      & $\NumDistillKLWeightedPair{}$ \\

    \midrule
    \multicolumn{5}{l}{\textit{Distillation candidate source}} \\
    Rollouts + impressions$^{\dagger}$
      & $\NumDistillBothRecallThousand{}$
      & $\mathbf{\NumDistillBothTeacherRecallTen{}}$
      & $\mathbf{\NumDistillBothTeacherRecallHundred{}}$
      & $\mathbf{\NumDistillBothWeightedPair{}}$ \\
    Rollouts only
      & $\NumDistillRolloutRecallThousand{}$
      & $\NumDistillRolloutTeacherRecallTen{}$
      & $\NumDistillRolloutTeacherRecallHundred{}$
      & $\NumDistillRolloutWeightedPair{}$ \\
    Impressions only
      & $\mathbf{\NumDistillImpressionRecallThousand{}}$
      & $\NumDistillImpressionTeacherRecallTen{}$
      & $\NumDistillImpressionTeacherRecallHundred{}$
      & $\NumDistillImpressionWeightedPair{}$ \\

    \midrule
    \multicolumn{5}{l}{\textit{Training rollout beam size}} \\
    32$^{\dagger}$
      & $\NumRolloutBeamThirtyTwoRecallThousand{}$
      & $\NumRolloutBeamThirtyTwoTeacherRecallTen{}$
      & $\NumRolloutBeamThirtyTwoTeacherRecallHundred{}$
      & $\NumRolloutBeamThirtyTwoWeightedPair{}$ \\
    64
      & $\NumRolloutBeamSixtyFourRecallThousand{}$
      & $\NumRolloutBeamSixtyFourTeacherRecallTen{}$
      & $\NumRolloutBeamSixtyFourTeacherRecallHundred{}$
      & $\NumRolloutBeamSixtyFourWeightedPair{}$ \\
    128
      & $\mathbf{\NumRolloutBeamOneTwentyEightRecallThousand{}}$
      & $\mathbf{\NumRolloutBeamOneTwentyEightTeacherRecallTen{}}$
      & $\mathbf{\NumRolloutBeamOneTwentyEightTeacherRecallHundred{}}$
      & $\mathbf{\NumRolloutBeamOneTwentyEightWeightedPair{}}$ \\
    \bottomrule
  \end{tabular}
  \caption{One-factor unified-model ablations around the shared reference:
  MAE, rollouts plus impressions, training rollout beam $32$,
  \NumBaselineHistoryLength{}-event history,
  \(E=\NumBaselineEncoderLayers{}\), \(D=\NumBaselineDecoderLayers{}\), and
  \(R=\NumItemScorerOneLayerDepth{}\). The daggered entries are the same
  reference run.}
  \label{tab:eval:unified-ablations}
\end{table}

\paragraph{Distillation loss.} We compare MAE, MSE, Huber, and pairwise KL on a
held-out split (\Cref{tab:eval:unified-ablations}). No loss dominates across
retrieval, ranking fidelity, and engagement-based ordering, so we retain MAE
as the reference for the subsequent ablations.

\paragraph{Distillation candidate distribution.}
Rollouts come from the current decoder; impressions come from logged requests.
Impressions alone have the highest retrieval point estimate but much lower
Teacher Recall than either configuration containing rollouts. Combining
both sources keeps Teacher Recall close to rollouts only and has the higher
WPA point estimate, so we use both sources.

\paragraph{Training rollout beam size.} The training beam
\(B_{\mathrm{train}}\) determines how many decoder candidates the Ranking
Module sees per training prompt; it is distinct from the fixed evaluation beam
\(B_{\mathrm{eval}}\). Increasing \(B_{\mathrm{train}}\) from 32 to 128 gives
the highest Teacher Recall point estimates but quadruples the number of rollout
candidates scored during training. We retain \(B_{\mathrm{train}}=32\) as the
lower-cost reference (\Cref{tab:eval:unified-ablations}).

\begin{table}[H]
  \centering
  \footnotesize
  \setlength{\tabcolsep}{2pt}
  \begin{tabular}{@{}llccccccc@{}}
    \toprule
    \multicolumn{5}{c}{\textbf{Configuration}}
      & \multicolumn{4}{c}{\textbf{Metrics}} \\
    \cmidrule(lr){1-5}
    \cmidrule(lr){6-9}
    \textbf{\shortstack{History\\encoding}}
      & \textbf{\shortstack{History\\Length}}
      & \textbf{\(E\)}
      & \textbf{\(D\)}
      & \textbf{\(R\)}
      & \textbf{\shortstack{Target-track\\Recall@1000}}
      & \textbf{\shortstack{Teacher\\Recall@10}}
      & \textbf{\shortstack{Teacher\\Recall@100}}
      & \textbf{WPA} \\
    \midrule
    Full attention
      & \NumBaselineHistoryLength{}
      & \NumBaselineEncoderLayers{}
      & \NumBaselineDecoderLayers{}
      & --
      & $\NumItemScorerNoneRecallThousand{}$
      & $\NumItemScorerNoneTeacherRecallTen{}$
      & $\NumItemScorerNoneTeacherRecallHundred{}$
      & $\NumItemScorerNoneWeightedPair{}$ \\
    Full attention
      & \NumBaselineHistoryLength{}
      & \NumBaselineEncoderLayers{}
      & \NumBaselineDecoderLayers{}
      & \NumItemScorerOneLayerDepth{}
      & $\NumItemScorerOneRecallThousand{}$
      & $\NumItemScorerOneTeacherRecallTen{}$
      & $\NumItemScorerOneTeacherRecallHundred{}$
      & $\NumItemScorerOneWeightedPair{}$ \\
    Full attention
      & \NumBaselineHistoryLength{}
      & \NumBaselineEncoderLayers{}
      & \NumBaselineDecoderLayers{}
      & \NumBaselineRankingModuleLayers{}
      & $\NumItemScorerFourRecallThousand{}$
      & $\NumItemScorerFourTeacherRecallTen{}$
      & $\NumItemScorerFourTeacherRecallHundred{}$
      & $\NumItemScorerFourWeightedPair{}$ \\
    \midrule
    Full attention
      & \NumScalingLongHistoryLength{}
      & \NumBaselineEncoderLayers{}
      & \NumBaselineDecoderLayers{}
      & \NumBaselineRankingModuleLayers{}
      & $\mathbf{\NumHistoryFullLongRecallThousand{}}$
      & $\mathbf{\NumHistoryFullLongTeacherRecallTen{}}$
      & $\mathbf{\NumHistoryFullLongTeacherRecallHundred{}}$
      & $\NumHistoryFullLongWeightedPair{}$ \\
    History Compression
      & \NumScalingLongHistoryLength{}
      & \NumBaselineEncoderLayers{}
      & \NumBaselineDecoderLayers{}
      & \NumBaselineRankingModuleLayers{}
      & $\NumHistoryCompressionRecallThousand{}$
      & $\NumHistoryCompressionTeacherRecallTen{}$
      & $\NumHistoryCompressionTeacherRecallHundred{}$
      & $\mathbf{\NumHistoryCompressionWeightedPair{}}$ \\
    \bottomrule
  \end{tabular}
  \caption{Ranking Module depth followed by history-length and
  history-encoding ablations. The \(E\), \(D\), and \(R\) columns report
  layer counts separately; \(R=\text{--}\) denotes decoder-likelihood
  ordering without a Ranking Module. The depth study increases \(R\) from
  \NumItemScorerOneLayerDepth{} to \NumBaselineRankingModuleLayers{}, and the
  history study fixes the latter depth.}
  \label{tab:eval:scorer-history}
\end{table}

\paragraph{Ranking Module depth and history encoding.}
\label{sec:eval:ranking}
\label{sec:eval:histcompress}

\Cref{tab:eval:scorer-history} first scales the Ranking Module with
\NumBaselineHistoryLength{}-event histories. Moving from decoder-likelihood
ordering (\(R=\text{--}\)) to one scorer layer substantially raises Teacher
Recall and WPA; increasing \(R\) to \NumBaselineRankingModuleLayers{} raises
the Teacher Recall point estimate further.

With \(R=\NumBaselineRankingModuleLayers{}\) fixed, extending full attention to
\NumScalingLongHistoryLength{} events gives the highest Target-track Recall and
Teacher Recall point estimates. History Compression
(\Cref{sec:arch:encoder,fig:sona-histcompress}) has close values and the highest
WPA point estimate while applying the deep encoder only to recent events. We
select this configuration for the final model.

\subsection{Online A/B Tests}
\label{sec:eval:online}

Unless marked \textsuperscript{\dag}, reported treatment--control deltas are
significant at \(p<0.01\). Five controlled A/B tests evaluate serving
configurations on My Vibe traffic from smart speakers. Experiments~1--3 examine
teacher-assisted configurations, Experiment~4 evaluates the distilled serving
model, and Experiment~5 evaluates the complete \Sona{} system.

Active Users is the primary metric; we also report Total Listening Time,
Likes, ``Repeat'' Commands, and Deeply Engaged Users. Every table entry is a
relative percentage change from the production control arm in the same
experiment.

\paragraph{Experiment 1: Likes score.} An NTP-only
encoder--decoder generates candidates, which the teacher orders using its
pairwise score plus a weighted likes score. We compare the
production control with three teacher-assisted treatments: pairwise scoring
alone (weight \(0\)), and likes-score weights \(\alpha\) and \(2\alpha\). Each
group contains \NumABExpLikeTrafficShare{} of users. The teacher attends to
\NumRewardHistoryLength{} events and the encoder--decoder to
\NumBaselineHistoryLength{}; its serving cost limits traffic and statistical
power (\Cref{tab:eval:ab:likehead}).

\begin{table}[H]
  \centering
  \small
  \setlength{\tabcolsep}{4pt}
  \begin{tabular}{lccc}
    \toprule
    \textbf{Metric}       & \textbf{Pairwise only (0)}
                          & \textbf{+ $\alpha$ Likes score}
                          & \textbf{+ $2\alpha$ Likes score} \\
    \midrule
    Total Listening Time  & $+\NumABPairwiseListeningTime\%$
                          & $+\NumABLikeOneListeningTime\%$
                          & $+\NumABLikeTwoListeningTime\%$ \\
    Likes                 & $\NumABPairwiseLikes\%$\textsuperscript{\dag}
                          & $+\NumABLikeOneLikes\%$
                          & $+\NumABLikeTwoLikes\%$ \\
    ``Repeat'' Commands   & $+\NumABPairwiseRepeats\%$
                          & $+\NumABLikeOneRepeats\%$
                          & $+\NumABLikeTwoRepeats\%$ \\
    Active Users          & $+\NumABPairwiseActiveUsers\%$
                          & $+\NumABLikeOneActiveUsers\%$
                          & $+\NumABLikeTwoActiveUsers\%$ \\
    Deeply Engaged Users  & $+\NumABPairwiseDeepUsers\%$
                          & $+\NumABLikeOneDeepUsers\%$
                          & $+\NumABLikeTwoDeepUsers\%$ \\
    \bottomrule
  \end{tabular}
  \caption{Experiment 1: control-relative changes for Likes-score weights in
  the teacher's ranking score, on the NTP-only encoder--decoder stack.}
  \label{tab:eval:ab:likehead}
\end{table}

The results support retaining a non-zero likes score in the teacher.

\paragraph{Experiment 2: Semantic ID features in the Teacher Ranker.}
Both treatments use the same NTP-only encoder--decoder for candidate generation
and the teacher for ordering. The ablation changes only the teacher's candidate
features: one teacher is trained without Semantic ID (SID) features, while the
other includes them. Control and both treatments each contain
\NumABExpSidTrafficShare{} of users (\Cref{tab:eval:ab:sidfeat}).

\begin{table}[H]
  \centering
  \small
  \begin{tabular}{lcc}
    \toprule
    \textbf{Metric}       & \textbf{Teacher without SID features}
                          & \textbf{Teacher with SID features} \\
    \midrule
    Total Listening Time  & $+\NumABNoSidListeningTime\%$
                          & $+\NumABSidListeningTime\%$ \\
    Likes                 & $+\NumABNoSidLikes\%$
                          & $+\NumABSidLikes\%$ \\
    ``Repeat'' Commands   & $+\NumABNoSidRepeats\%$
                          & $+\NumABSidRepeats\%$ \\
    Active Users          & $+\NumABNoSidActiveUsers\%$
                          & $+\NumABSidActiveUsers\%$ \\
    Deeply Engaged Users  & $+\NumABNoSidDeepUsers\%$
                          & $+\NumABSidDeepUsers\%$ \\
    \bottomrule
  \end{tabular}
  \caption{Experiment 2: control-relative changes when the teacher's SID
  candidate features are the only difference between treatments; both use the
  same NTP-only encoder--decoder.}
  \label{tab:eval:ab:sidfeat}
\end{table}

We therefore retain SID features in the teacher.

\paragraph{Experiment 3: Teacher Ranker and candidate generation.} We test the
teacher with two candidate-generation paths. One treatment replaces only the
production ranker; the other also replaces production candidate generation
with the encoder--decoder. Both use only the teacher's pairwise score, matching
the scoring objective across treatments. Serving cost limits each split to
\NumABExpTeacherTrafficShare{} of users
(\Cref{tab:eval:ab:teacher}).

\begin{table}[H]
  \centering
  \small
  \begin{tabular}{lcc}
    \toprule
    \textbf{Metric}         & \textbf{Teacher Ranker}
                            & \textbf{Enc--dec + Teacher Ranker} \\
    \midrule
    Total Listening Time    & $+\NumABTeacherListeningTime\%$
                            & $+\NumABStackListeningTime\%$ \\
    Likes                   & $+\NumABTeacherLikes\%$\textsuperscript{\dag}
                            & $\NumABStackLikes\%$\textsuperscript{\dag} \\
    ``Repeat'' Commands     & $+\NumABTeacherRepeats\%$\textsuperscript{\dag}
                            & $+\NumABStackRepeats\%$\textsuperscript{\dag} \\
    Active Users            & $+\NumABTeacherActiveUsers\%$
                            & $+\NumABStackActiveUsers\%$ \\
    Deeply Engaged Users    & $+\NumABTeacherDeepUsers\%$
                            & $+\NumABStackDeepUsers\%$ \\
    \bottomrule
  \end{tabular}
  \caption{Experiment 3: the Teacher Ranker replacing the production
  ranker, and the NTP-only encoder--decoder with Teacher Ranker
  ordering replacing the full stack.}
  \label{tab:eval:ab:teacher}
\end{table}

The Teacher Ranker can successfully replace the production ranker, and
the further replacement of candidate generation by the encoder--decoder
shows promising results toward replacing the full stack.

\paragraph{Experiment 4: distillation.} We compare the distilled model with the
same candidate generator re-ranked by the teacher. Both treatments use the
encoder--decoder trained with the joint objective of
\Cref{sec:train:pretrain} and \NumBaselineHistoryLength{}-event histories. One
orders candidates with the distilled Ranking Module; the other uses the
Teacher Ranker, which attends to \NumRewardHistoryLength{} events. Both include
the likes score. Control and both treatments each contain
\NumABExpDistillTrafficShare{} of users (\Cref{tab:eval:ab:distill}).

\begin{table}[H]
  \centering
  \small
  \begin{tabular}{lcc}
    \toprule
    \textbf{Metric}         & \textbf{Distilled model}
                            & \textbf{Distilled + Teacher Ranker} \\
    \midrule
    Total Listening Time    & $+\NumABDistilledListeningTime\%$
                            & $+\NumABDistilledTeacherListeningTime\%$ \\
    Likes                   & $+\NumABDistilledLikes\%$
                            & $+\NumABDistilledTeacherLikes\%$ \\
    ``Repeat'' Commands     & $+\NumABDistilledRepeats\%$
                            & $+\NumABDistilledTeacherRepeats\%$ \\
    Active Users            & $+\NumABDistilledActiveUsers\%$
                            & $+\NumABDistilledTeacherActiveUsers\%$ \\
    Deeply Engaged Users    & $+\NumABDistilledDeepUsers\%$
                            & $+\NumABDistilledTeacherDeepUsers\%$ \\
    \bottomrule
  \end{tabular}
  \caption{Experiment 4: the distilled single model, alone and
  re-ranked by the Teacher Ranker, replacing the full stack.}
  \label{tab:eval:ab:distill}
\end{table}

The distilled-model result shows that the production cascade can be replaced
without serving the teacher.

\paragraph{Experiment 5: \Sona{}.} The fifth experiment studies the
further improvement of the recipe by extending the user history to
\NumCompressMaxLen{} events with History Compression. It has two
splits --- control and \Sona{}, the full recipe described in this report
--- and ran for \NumABSonaDays{} days on \NumABSonaTrafficShare{} of
randomly selected users in each split. Hard business rules that
restrict production output are preserved (\Cref{tab:eval:ab:sona}).

\begin{table}[H]
  \centering
  \small
  \begin{tabular}{lc}
    \toprule
    \textbf{Metric}         & \textbf{Relative change vs control} \\
    \midrule
    Total Listening Time    & $+\NumABSonaListeningTime\%$ \\
    Likes                   & $+\NumABSonaLikes\%$ \\
    ``Repeat'' Commands     & $+\NumABSonaRepeats\%$ \\
    Active Users            & $+\NumABSonaActiveUsers\%$ \\
    Deeply Engaged Users    & $+\NumABSonaDeepUsers\%$ \\
    \bottomrule
  \end{tabular}
  \caption{Experiment 5: the single \Sona{} model replaces the whole
  recommendation stack.}
  \label{tab:eval:ab:sona}
\end{table}

A key difference from Experiment~4 is the available user context:
Experiment~4 uses \NumBaselineHistoryLength{}-event histories, whereas
Experiment~5 extends the history to \NumCompressMaxLen{} events with History
Compression. The offline history ablation
(\Cref{tab:eval:scorer-history}) shows material gains from the longer context,
so this change is an important part of the complete \Sona{} configuration.
Because the two online experiments were conducted separately, their difference
does not isolate the causal contribution of history length alone.

On the primary Active Users metric, \Sona{}'s
$+\NumABSonaActiveUsers{}\%$ control-relative uplift is
\NumABVsArgusRatio{} as large as the $+\NumABArgusActiveUsers{}\%$ uplift
previously recorded for Argus \citep{argus2025}, the strongest model previously
deployed on this surface.

\section{Conclusion, Limitations, and Future Work}
\label{sec:conclusion}

This technical report demonstrates that a mature industrial
recommendation stack --- candidate generation, pre-ranking, and
ranking --- can be replaced by a single model trained and served end
to end. \Sona{} generates candidates and ranks them with one
transformer over the user's raw event history, supervised by
next-token prediction and by distillation from a Teacher Ranker without
hand-engineered features. In its final online A/B test, it improves all five
reported engagement metrics relative to the production control
(\Cref{sec:eval:online}). The remainder of this section lists the
limitations of the present system and the directions we are pursuing
next.

\paragraph{Toward full deployment.} Despite the major improvements
in core platform metrics, we have not yet deployed \Sona{} on full
traffic. \Sona{} introduces significant changes to the existing
stack, and we want to validate it in a long-term experiment spanning
a multi-month window. We also observed that \Sona{}'s catalog
coverage is lower than that of the production stack, and we are going
to investigate this effect.

\paragraph{Other surfaces.} This report validates \Sona{} on one of
our recommendation surfaces. Full deployment requires validating the
approach on the other surfaces, and we leave this investigation for
future work.

\paragraph{Model scaling.} Across the measured capacity and data ranges, the
reported point estimates continue to improve at the largest tested settings
(\Cref{sec:eval:retrieval}). We will continue
enlarging the model's capacity, including sparse mixture-of-experts
backbones, the adoption of sparse and linear attention
\citep{gbla2026}, and a persistent cross-request KV cache.

\paragraph{Test-time scaling.} The potential of test-time scaling is
still underexplored. Beam search provides such an axis --- widening
the beam trades inference compute for candidate quality --- but how
to leverage test-time scaling properly remains an open question.

\paragraph{Content exploration.} We did not notice any degradation in
content exploration, which could be explained by \Sona{} being
trained on production logs. We leave further investigation of
fresh-content behavior for future work.

\paragraph{Reinforcement learning.}
Reinforcement-learning post-training has proved effective for LLMs
\citep{stiennon2020learning, nakano2021webgpt, ouyang2022training,
bai2022helpful, guo2025deepseek}, and industrial generative
recommenders increasingly use RL to align generation with engagement
and business objectives \citep{onerec2025, oxygenrec2025, qiu2025,
zheng2025, oneloc2025, onemall2026, onelive2026}. Recommendation is a
natural fit: logged user feedback supplies a verifiable reward, and
that reward is tightly coupled with the business objectives the
platform optimizes. We adopt a training stage only once a controlled
comparison demonstrates its contribution, and for RL we have not yet
built that evidence; the recipe reported here is therefore fully
supervised. Promising directions for \Sona{} include long-term reward
modeling and RLVR-style post-training on user feedback, and we leave
their controlled evaluation to future work.

\bibliographystyle{plainnat}
\bibliography{bibliography}

\appendix
\section{Contributions and Acknowledgments}
\label{app:contributions}

Contributors are listed alphabetically by first name.
\textsuperscript{*}~marks contributors who have since departed from the
team.

\paragraph{Contributors.}
Alexandr Udeneev,
Aleksei Krasilnikov\textsuperscript{*},
Alexey Nadtochiy,
Andrey Semenov,
Andrey Tsyrkunov,
Anna Krivonos,
Anna Lipkina,
Artem Matveev\textsuperscript{*},
Daniil Burlakov,
Daniil Leshchev\textsuperscript{*},
Daria Tikhonovich,
Denis Burshtein,
Ekaterina Dmitrieva,
Eugene Krofto,
Grigorii Khlystov\textsuperscript{*},
Ilya Murzin,
Kirill Golovko,
Ksenia Sycheva,
Leonid Dmitriev,
Mariia Rozaeva,
Mariia Ulianova,
Mikhail Sandul,
Nikolai Savushkin,
Oleg Sorokin,
Roman Odobesku,
Semyon Panenko,
Sergei Liamaev\textsuperscript{*},
Sergei Makeev\textsuperscript{*},
Vadim Shilov,
Veronika Ivanova,
Viktor Yanush,
Vladimir Baikalov\textsuperscript{*},
Vladislav Dodonov,
Vladislav Tytskiy\textsuperscript{*}.

\paragraph{Acknowledgments.}
We would like to extend special thanks to
Alexander Ploshkin,
Alexey Gusakov,
Alexey Pismenny,
Nikita Stepanov,
Petr Ermakov,
Petr Garmider,
and Sergey Kastryulin
for their substantial support throughout this work.

\section{Hyperparameters and Model Cards}
\label{app:hyperparameters}

\addtocontents{toc}{\protect\setcounter{tocdepth}{0}}

\subsection{Semantic Tokenizer}

\begin{table}[H]
  \centering
  \small
  \begin{tabularx}{0.85\linewidth}{lX}
    \toprule
    \textbf{Parameter} & \textbf{Value} \\
    \midrule
    Audio foundation model           & Qwen2.5-Omni \citep{qwen25omni} \\
    Audio input length               & 90 s mel-spectrogram \\
    Embedding refresh threshold      & \NumPopThreshold{} lifetime plays \\
    Qwen output sequence length      & up to \NumQwenSeqLen{} hidden vectors \\
    Refinement transformer           & 4 layers, 4 heads, model dim 512 \\
    Item embedding dimension         & \NumTokenizerEmbDim{} \\
    Refinement contrastive temperature & 0.1 \\
    Refinement alignment weight      & 0.1 \\
    Semantic ID codebooks            & \NumSemIDLevels{} levels,
                                       \NumCodebookSize{} codes per level \\
    \bottomrule
  \end{tabularx}
  \caption{Tokenizer model card.}
  \label{tab:hparams:tokenizer}
\end{table}

\subsection{Teacher Ranker}

\begin{table}[H]
  \centering
  \small
  \begin{tabularx}{0.85\linewidth}{lX}
    \toprule
    \textbf{Parameter} & \textbf{Value} \\
    \midrule
    Encoder layers                   & \NumRewardEncoderLayers{} (causal) \\
    Candidate-scorer layers          & \NumRewardCrossLayers{} \\
    Hidden dimension                 & \NumRewardHidden{} \\
    Attention heads                  & \NumRewardHeads{} (head dim 64) \\
    Hash embedding tables            & $2^{19} \times 256$ (item / artist),
                                       $2^{19} \times 512$ (semantic prefix) \\
    Per-level SID embeddings         & $32001 \times 128$ \\
    Maximum positions                & 8193 \\
    Dropout                          & 0.1 (encoder); 0 (candidate scorer) \\
    Total parameters                 & \NumRewardParams{} (embedding-table dominated) \\
    Pre-training history length      & \NumRewardPretrainLength{} events \\
    Fine-tuning history length       & \NumRewardHistoryLength{} events \\
    NIP in-batch negatives           & $2^{13} = 8192$ \\
    LR schedule                      & $10^{-5} \to 10^{-4}$ over 3000 warmup
                                       steps, then constant (both groups) \\
    Gradient clipping                & 1.0 \\
    \bottomrule
  \end{tabularx}
  \caption{Teacher Ranker card. Initial training has two stages
  (\Cref{sec:teacher:training}); only the encoder is reused between
  pre-training and multi-head ranking fine-tuning. The resulting model is
  then refreshed continuously.}
  \label{tab:hparams:reward}
\end{table}

\subsection{Ranking Module}

\begin{table}[H]
  \centering
  \small
  \begin{tabularx}{0.85\linewidth}{lX}
    \toprule
    \textbf{Parameter} & \textbf{Value} \\
    \midrule
    Blocks                           & \NumBaselineRankingModuleLayers{}
                                       cross-attention layers \\
    Hidden dimension                 & \NumHiddenDim{} \\
    Attention heads                  & 4 (head dim 256) \\
    Feed-forward dimension           & 4096 \\
    Scoring outputs                  & 2 teacher-head scores \\
    Rollouts per prompt              & 32 \\
    Loss weights                     & NTP : rollout : impression $= 1:1:1$ \\
    \bottomrule
  \end{tabularx}
  \caption{Ranking Module card. Distilled from the Teacher Ranker
  jointly with encoder--decoder backbone training (\Cref{sec:train:rescorer}).}
  \label{tab:hparams:rescorer}
\end{table}

\subsection{Encoder--Decoder Backbone}

\begin{table}[H]
  \centering
  \small
  \begin{tabularx}{0.85\linewidth}{lX}
    \toprule
    \textbf{Parameter} & \textbf{Value} \\
    \midrule
    Layer allocation                 & \NumBackboneLayers{} recent-block layers;
                                       \NumFullHistoryLayers{} full-history layer;
                                       \NumBridgeLayers{} bridge per direction \\
    Hidden dimension                 & \NumHiddenDim{} \\
    Attention heads                  & \NumAttnHeads{} (head dim 64) \\
    Feed-forward dimension           & 2816 \\
    History lengths                  & \NumCompressMaxLen{} events total;
                                       \NumCompressRecentLen{} recent
                                       (+1 \textsc{cls}) \\
    Hash-bucket embedding table      & $2^{19} \times 256$ (shared) \\
    \bottomrule
  \end{tabularx}
  \caption{Encoder model card.}
  \label{tab:hparams:encoder}
\end{table}

\begin{table}[H]
  \centering
  \small
  \begin{tabularx}{0.85\linewidth}{lX}
    \toprule
    \textbf{Parameter} & \textbf{Value} \\
    \midrule
    Layers                           & \NumDecoderLayers{} (causal
                                       self-attention + cross-attention
                                       over encoder states) \\
    Semantic ID vocabulary           & \NumSemIDLevels{} codebooks
                                       $\times$ \NumCodebookSize{}
                                       (+ \textsc{bos} row) \\
    Output heads                     & \NumSemIDLevels{} $\times$
                                       Linear($1024 \to 32000$), untied \\
    Serving decoding                 & constrained beam search, beam
                                       \NumInferenceBeamSize{} \\
    \bottomrule
  \end{tabularx}
  \caption{Decoder model card.}
  \label{tab:hparams:decoder}
\end{table}

\begin{table}[H]
  \centering
  \small
  \begin{tabularx}{0.85\linewidth}{lX}
    \toprule
    \textbf{Parameter} & \textbf{Value} \\
    \midrule
    Optimizer                        & AdamW (fused), weight decay $0.1$,
                                       gradient clipping $1.0$ \\
    LR schedule (initial training)   & linear, $10^{-5} \to 3\times10^{-4}
                                       \to 7\times10^{-5}$,
                                       3000-step warmup \\
    Effective batch size             & $16{,}384$ samples \\
    \bottomrule
  \end{tabularx}
  \caption{Encoder--decoder backbone optimization.}
  \label{tab:hparams:backbone}
\end{table}

\subsection{Online Training and Model Refresh}
\label{app:hparams:online}

\begin{table}[H]
  \centering
  \small
  \begin{tabularx}{0.85\linewidth}{lX}
    \toprule
    \textbf{Configuration} & \textbf{Value} \\
    \midrule
    Online update mode
      & continuous optimization over the event-ordered request stream \\
    LR schedule
      & constant $7\!\times\!10^{-5}$, the terminal rate of the
        initial-training schedule \\
    Training objective
      & next-token prediction and Ranking Module distillation \\
    Session-attribution window
      & \NumAttributionWindow{} \\
    Serving-model sync cadence
      & every \NumModelSyncPeriod{} \\
    Teacher snapshot refresh cadence
      & every \NumRewardRefreshCadence{} \\
    \bottomrule
  \end{tabularx}
  \caption{Online-training and model-refresh configuration
  (\Cref{sec:infra:online}).}
  \label{tab:hparams:online}
\end{table}

\addtocontents{toc}{\protect\setcounter{tocdepth}{2}}

\section{Notation}
\label{app:notation}

\begin{table}[H]
  \centering
  \small
  \begin{tabularx}{\linewidth}{lX}
    \toprule
    \textbf{Symbol} & \textbf{Definition} \\
    \midrule
    \multicolumn{2}{l}{\emph{User history and encoder}} \\
    $u = (e_1, \ldots, e_T)$ & chronological user engagement history;
      $e_t$ is one logged event \\
    $T$, $T_{\max}$  & history length and its cap \\
    $X$              & embedded event sequence, the encoder input
      (\Cref{sec:arch:history}) \\
    $H = \mathrm{Enc}_{\theta}(X)$ & direct hidden-state output of a
      full encoder, $H \in \mathbb{R}^{(T+1) \times d}$ \\
    $d$              & hidden dimension \\
    $\theta$         & learnable model parameters \\
    \midrule
    \multicolumn{2}{l}{\emph{Semantic tokenizer}} \\
    $v$              & catalog track \\
    $s(v) = (s_1, \ldots, s_L)$ & Semantic ID tuple of track $v$;
      $s_\ell$ is the level-$\ell$ code, $L$ the number of levels \\
    $z_a$, $\bar h_a$ & item embedding of track $a$ and the mean of
      its content features \\
    $m$              & collaborative pairs per training batch \\
    $\mathrm{sim}$, $\tau$ & cosine similarity and InfoNCE temperature \\
    $\lambda_{\mathrm{align}}$ & weight of the alignment loss term \\
    $\mathrm{NPMI}(i, j)$ & normalized pointwise mutual information of
      co-listening tracks $i$ and $j$; $p(i, j)$, $p(i)$ are
      co-occurrence and marginal probabilities \\
    \midrule
    \multicolumn{2}{l}{\emph{History Compression}} \\
    $N$              & maximum attended history length \\
    $O$, $R$         & long-term and recent blocks of
      $n_{\mathrm{o}} = N - n_{\mathrm{r}}$ and $n_{\mathrm{r}}$
      events \\
    $\tilde O$, $\tilde R$ & blocks enriched by bidirectional
      cross-attention \\
    $X_O$, $X_R$     & block representations after the full-history
      pass \\
    $H_R$            & deep-stack encoding of the recent block $X_R$ \\
    $K$              & shared encoder memory exposed to the decoder and
      Ranking Module; $K=H$ for a full encoder and
      $K=[\,X_O;H_R\,]$ with History Compression \\
    \midrule
    \multicolumn{2}{l}{\emph{Training}} \\
    $C^{+}(u)$       & packed positive target set of the next-token
      objective \\
    $B$              & rollout beam size \\
    $\mathcal{B}$, $\mathcal{I}$ & rollout candidate set and logged
      impressions of a request, the two distillation sources \\
    $r^{h}_{c}$, $\hat r^{\,h}_{c}$ & teacher and Ranking Module
      head-$h$ scores of candidate $c$; $n$ is the number of distilled
      heads \\
    $\mathcal{L}_\text{NTP}$, $\mathcal{L}_\text{rollout}$,
      $\mathcal{L}_\text{impression}$ & next-token loss and the two
      distillation losses of the joint objective
      (\Cref{sec:train:pretrain}) \\
    \midrule
    \multicolumn{2}{l}{\emph{Teacher Ranker}} \\
    $\mathcal{P}$    & ordered pairs of temporally adjacent
      impressions with different engagement grades \\
    $\mathrm{score}_{\mathrm{pair}}(i)$ & pairwise-head score of
      impression $i$ \\
    $\mathcal{L}_\text{pair}$, $\mathcal{L}_{\mathrm{BCE}}^{k}$,
      $\mathcal{L}_\text{teacher}$ & pairwise loss, per-head pointwise
      losses, and their sum (\Cref{sec:teacher:training}) \\
    \midrule
    \multicolumn{2}{l}{\emph{Evaluation}} \\
    $k$              & ranking cutoff \\
    $\mathrm{Top}_k(\cdot)$ & top-$k$ item set under a scoring
      function \\
    $r$, $\mathcal{R}$ & evaluated request/history and the set of such
      requests \\
    $C_r$            & decoder-generated candidate pool for request $r$ \\
    $s_{\mathrm{RM}}^{(r)}$, $s_{\mathrm{teacher}}^{(r)}$ &
      Ranking Module and teacher scores over $C_r$ \\
    $t_i$            & predefined target weight of impression $i$'s
      engagement label \\
    \bottomrule
  \end{tabularx}
  \caption{Notation used throughout the report.}
  \label{tab:notation}
\end{table}

\end{document}